\documentclass[]{aastex63}
\usepackage[caption=false]{subfig}
\usepackage{float}
\usepackage{multirow}

\shorttitle{TDA and Solar Flare Prediction}
\shortauthors{Deshmukh et al.}
\graphicspath{{./}{figures/}}

\begin{document}

\title{Machine Learning Approaches to Solar-Flare Forecasting: Is
  Complex Better?}

\correspondingauthor{Varad Deshmukh}
\email{varad.deshmukh@colorado.edu}

\author[0000-0003-1858-8044]{Varad Deshmukh}
\affiliation{Department of Computer Science \\ 
University of Colorado Boulder \\
Boulder CO 80309-0430 USA}

\author{Srinivas Baskar}
\affiliation{Department of Computer Science \\ 
University of Colorado Boulder \\
Boulder CO 80309-0430 USA}


\author[0000-0002-4567-2543]{Elizabeth Bradley}
\affiliation{Department of Computer Science \\ 
University of Colorado Boulder \\
Boulder CO 80309-0430 USA}
\affiliation{The Santa Fe Institute \\ 
Sante Fe NM 87501 USA}

\author[0000-0002-4989-475X]{Thomas Berger}
\affiliation{Space Weather Technology Research and Education Center\\
Boulder CO 80309-0429 USA}

\author[0000-0002-0019-0356]{James D. Meiss}
\affiliation{Department of Applied Mathematics \\ 
University of Colorado Boulder \\
Boulder CO 80309-0526 USA}



\begin{abstract}

Recently, there has been growing interest in the use of
machine-learning methods for predicting solar flares.  Initial efforts
along these lines employed comparatively simple models, correlating
features extracted from observations of sunspot active regions
with known instances of flaring. Typically, these models have used
physics-inspired features that have been carefully chosen by experts
in order to capture the salient features of such magnetic field
structures.  Over time, the sophistication
and complexity of the models involved has grown.  However, there has
been little evolution in the choice of feature sets, 
nor any systematic study of whether the additional model complexity is
truly useful.  Our goal is to address these issues.  To
that end, we compare the relative prediction performance of
machine-learning-based, flare-forecasting models with varying degrees
of complexity.  We also revisit the feature set design, using
topological data analysis to extract shape-based features
from magnetic field images of the active regions. Using 
hyperparameter training for fair comparison of different
machine-learning models across different feature sets, we show that
simpler models with fewer free parameters \textit{generally perform better
than more-complicated models}, ie., powerful machinery does not 
necessarily guarantee better prediction performance. Secondly,  we find that 
\textit{abstract, shape-based features contain just as much useful
  information}, for the purposes of flare prediction, as the
set of hand-crafted features developed by the solar-physics community
over the years.
Finally, we study the effects of dimensionality reduction, using
principal component analysis, to show that streamlined feature sets,
overall, perform just as well as the corresponding full-dimensional versions.

\end{abstract}



\section{Introduction} \label{sec:intro}

Solar flares are produced as a result of intense release of magnetic
energy from active regions
on the surface of the Sun.  The high-energy plasma emitted in
conjunction with these eruptions can cause catastrophic events on
Earth, with potentially trillions of dollars in associated economic
losses.  With enough notice, it is possible to mitigate some of the
effects of these events, so forecasting is of obvious importance.  In
current operational practice, flare forecasts are produced by human
experts using established classification systems
\citep{McIntosh:1990wu,Hale:1919} to categorize active regions into
various classes. A forecast is then constructed by consulting lookup
tables of flaring rates for each category, which are derived from
historical records \citep{Crown:2012}.  Over the past two decades,
there has been a great deal of effort devoted to machine learning (ML)
solutions to this problem, as discussed below.
After being trained on corpora of observations of the Sun to learn
correlations between the data and known instances of solar flares,
these models can be applied to new observations to generate flaring
forecasts.

The most commonly used observations in solar-flare forecasting are
magnetic field images called \textit{magnetograms} that are captured
by the Helioseismic and Magnetic Imager (HMI) onboard the Solar
Dynamics Observatory (SDO).  The SDO, located in an inclined 
geosynchronous orbit around Earth, has been
operational since 2010, recording the vector magnetic field data on
the side of Sun's photosphere that is visible from Earth.  HMI supplies not only
full-disk images of the Sun, but also cutouts of each active region as
it moves across the photosphere, recorded at a cadence of 12 minutes.
These cutouts are called \textit{Spaceweather HMI Active Region
  Patches} or \text{SHARPs}.  Each SHARPs record contains values for
the radial, polar, and azimuthal components $B_r$, $B_\theta$, and
$B_\phi$ of the magnetic field at the location sampled by each pixel,
along with a number of aggregate properties, such as the magnetic flux
or the electric current. These physics-inspired attributes are
carefully chosen by experts so as to be relevant to the flaring
phenomenon.

This information has been used in different ways in a variety of
machine-learning methods.  Some researchers have trained models
directly on the raw SHARPs images: e.g., \citep{Huang2018, Park2018,
  Zheng2019, Li2020, Abed2021, Zheng2021}.  However, it is well known in the ML
literature that ``featurizing'' data---preprocessing it
to extract higher-level properties that are salient in a given context---can
be extremely advantageous.  The notion of
salience is problem-specific; in computer vision, for instance, useful
features might be edges or polygons.  The attributes in the SHARPs
metadata are a natural feature set for the flare forecasting problem,
and most ML work to date has followed that reasoning.  A wide range of
models have been trained on this feature set: linear discriminant
analysis \citep{Leka2007}, logistic regression \citep{yuan2010}, LASSO
regression \citep{Campi2019}, support vector machines
\citep{bobra}, random forests or extremely randomized trees
\citep{Nishizuka:2017, Campi2019}---and, in recent years,
deep-learning models like multilayer perceptrons
\citep{Nishizuka:2018}, long short-term memories \citep{Chen2019}, and
autoencoders \citep{Chen2019}.

The performance of these ever-more-complicated models, however, is no
better than that of the human forecasters \citep{barnes_survey,
  leka2019a, leka2019b}.  Moreover,
high model complexity may not actually be an advantage in this
application. It is possible that this is due simply to data limitations: in
general, more-complex machine-learning models require larger training
sets to effectively extract the patterns needed for prediction.  This
issue is exacerbated when correlations are deeply embedded in complex
data for complex situations.  A second issue is the nature of
the training data.  Until very recently, ML-based flare forecasting
work had not moved beyond the original SHARPs data: that is, the
images themselves (i.e., the $B$ values at each pixel) and the
associated physics-based features.  Choosing features that are
scientifically meaningful makes good sense, of course, but ML methods
can sometimes leverage attributes that are not obvious to human experts.  For
example, a recent \textit{Nature} paper reports on using ML to
discover a previously unknown correlation between geometric and
topological attributes of knots \citep{davies2021}
A third and related concern is the dimensionality of the training
data.  This, too, is a well known issue in the ML community,
but there has been little exploration of this matter in the context of
ML-based flare forecasting methods.  Existing approaches have used the
\textit{complete} SHARPs feature set, but it may well be the case that
a carefully crafted subset of these features would work as well---or
perhaps even better.

In this paper, we aim to address the issues raised in the previous
paragraph.  Our first objective is to make a systematic, meaningful
comparison of the predictive performance of machine-learning
flare-forecasting models with varying complexity.  The specific
question we aim to address is: \textit{How does increasing model
  complexity affect 24-hour forecast accuracy for solar flares?}  To
that end, we use four ML models: logistic regression, extremely
randomized trees (ERTs), multilayer perceptrons (MLPs), and long
short-term memories (LSTMs).  Here, we define complexity informally as
the number of parameters that are optimized during the training of the
model. For example, logistic regression requires calculating a scaling
weight for each feature involved, whereas MLPs and LSTMs are complex
networks with hundreds of nonlinear, weighted connections that are
adjusted during training.  Comparing different ML models is not a
trivial task, given the complexity of the training process.  Each
method has a number of free parameters, known as
\textit{hyperparameters}, that control the learning process: the
learning rate in neural networks, for instance,
or the maximum number of trees in an ERT. In order to make a
systematic and fair comparison, we use an established hyperparameter
tuning framework \citep{liaw2018tune} to optimize the performance of
each model, then train it on a set of SHARPs from the period
2010-2017, each labeled as to whether that region produced a flare within
the following 24 hours.  Finally, we use various standard metrics to
compare the predictive accuracy of these models on a different set of
similarly labeled active regions.
The results show that the simpler models (logistic regression and ERT)
perform better than the more-complicated MLP and LSTM models.

The second objective of this paper is to explore \textit{whether there
  are useful ways to move beyond the set of physics-based attributes
  in the SHARPs metadata}.  Conjecturing that the shape of the
magnetic structures in the photosphere could provide important clues
about the evolving complexity of the field leading up to an eruption,
we focus specifically on the abstract spatial properties of the
magnetograms, such as the number of ``holes" in a thresholded image of
the magnetic field strength, extracted using topological data
analysis.  Our preliminary studies using multilayer perceptron models
\citep{Deshmukh2020, Deshmukh2020_IAAI} suggested that features like
this could perform as well as a basic set of SHARPs features.  This
was significant because the SHARPs feature set has been designed by
experts, while the topological features are straightforward spatial
attributes that require no hand crafting. However, that study used a
single ML model and the comparison only involved a subset of the
SHARPs features.  In this paper, we perform a more-extensive
comparison using the four ML models listed above and a
more-comprehensive physics-based feature set.
We also explore dimensionality reduction.  Feature sets often contain
correlations and redundancies: e.g., the mean and total values of some
derived quantity.  Removing these redundancies can both speed training
and increase accuracy.  Here, we use principal component analysis
and find that models trained on the most
significant principal components of each feature set\footnote{those
explaining 98.5\% of the variance} largely equaled the performance of
those trained on the full set of features.  This is an
advantage since the amount of data required to
successfully train a machine-learning model grows with its dimensionality.

The outline of this paper is as follows. Section~\ref{sec:data}
describes the details of the magnetogram data set and the labeling
procedure.  The different feature sets---SHARPs and topological---are
discussed in Section~\ref{sec:features}.  Section~\ref{sec:models}
describes the machine-learning models, 
the hyperparameter tuning framework, and the
dimensional reduction strategy.
The results are given in Section~\ref{sec:Results}, and
Section~\ref{sec:conclusion} concludes.

\section{Data} 
\label{sec:data}

\begin{figure}
	\begin{center} 
   \gridline{
   	   \fig{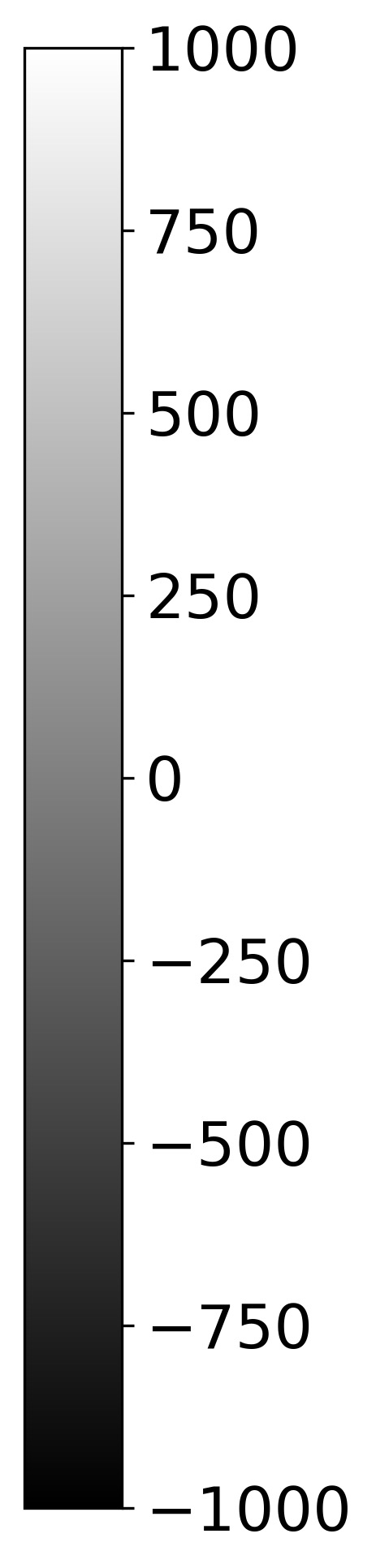}{0.052\textwidth}{}
   	   \fig{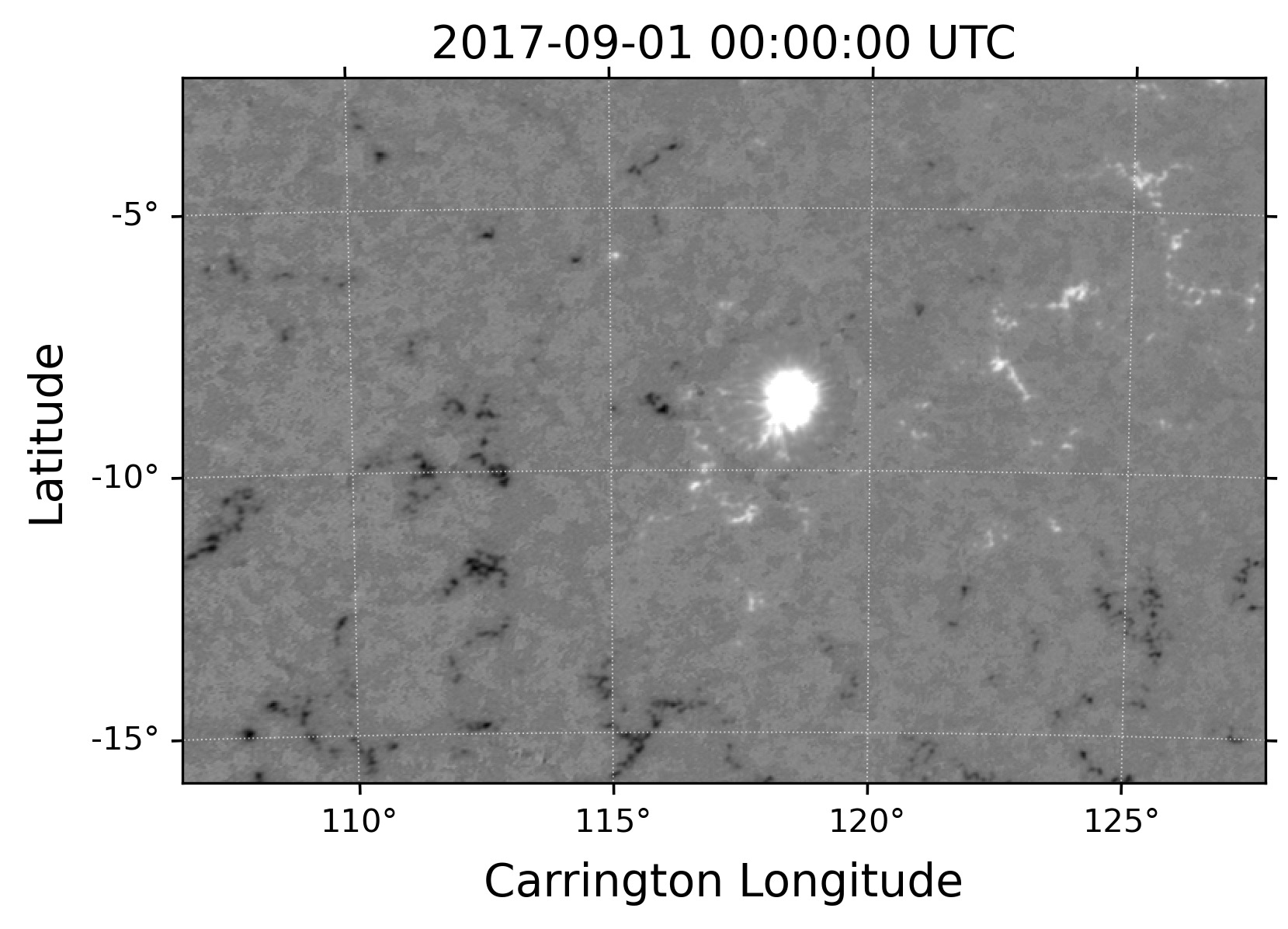}{0.30\textwidth}{(a)}
   	   \fig{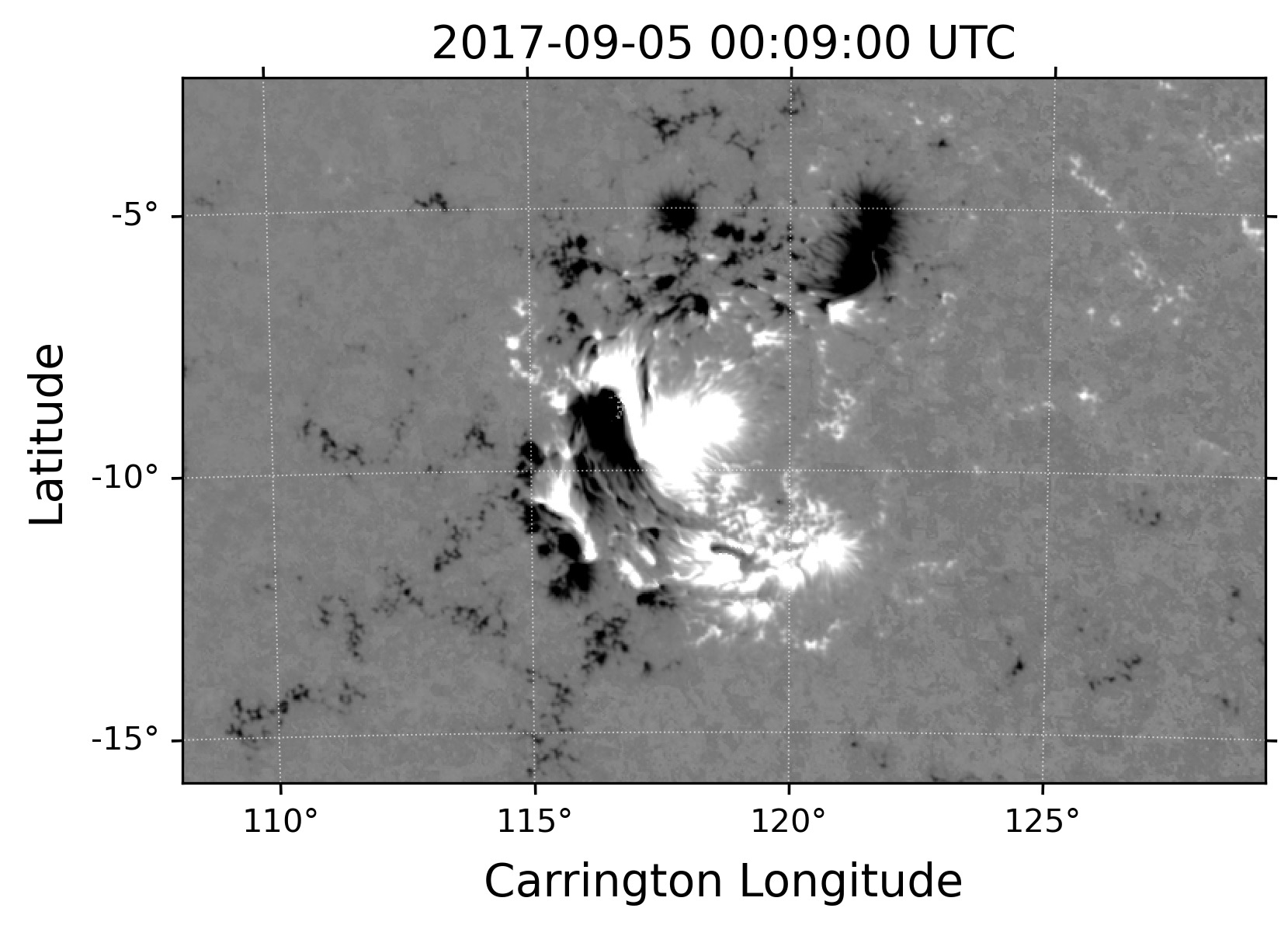}{0.30\textwidth}{(b)}
   	   \fig{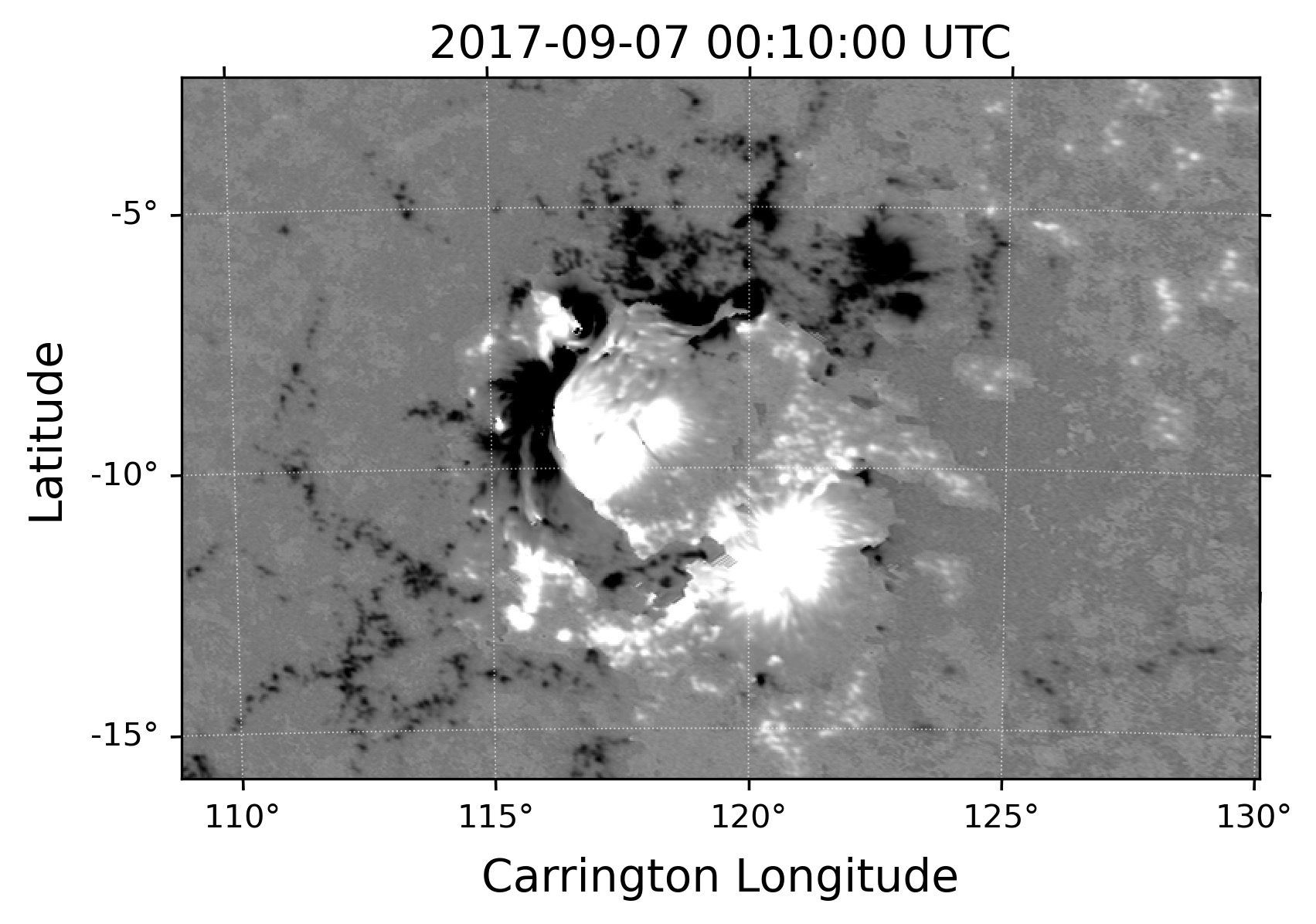}{0.30\textwidth}{(c)}
	}
                
   \end{center} 
   \caption{Radial component of magnetogram observations of sunspot
     \#AR 12673, which produced M- and X-class flares during its
     journey across the solar disk in September 2017: (a) at 0000 UT
     on 9/1, (b) at 0900 UT on 9/5, approximately 24 hours before
     producing an X-class flare, and (c) at 1000 UT on 9/7, during
     an M-class flare.}
	\label{fig:magnetograms}
\end{figure}

We use magnetic field images captured by the Helioseismic and Magnetic
Imager (HMI) instrument onboard the Solar Dynamics Observatory for our
study \citep{Scherrer2012}.  This data set is available on the
Stanford Joint Space Operations Center in the form of the Spaceweather
HMI Active Region Patches mentioned in Section~\ref{sec:intro}: rectangular cutouts of the
photospheric vector magnetic fields.  Here we use the {\tt
  hmi.sharp\_cea\_720s} variant, which stores the Lambert Cylindrical
Equal-Area projection of the magnetic field.  Of the three field components
in each SHARPs image,
we use only the radial component, $B_r$. 
This is standard practice in ML-based flare forecasting work, as the radial surface 
field is used as the boundary condition for computing global coronal magnetic 
fields that are not yet routinely measurable \citep{Caplan2021}.
Figure~\ref{fig:magnetograms} shows a series of SHARPs
observations of a single active region.
In each panel, white represents the positive polarity of the radial
flux, while black corresponds to negative polarity.  Magnetic loops
formed from emerging flux on the surface of the Sun manifest as pairs
of dark and bright spots that represent their footpoints.  Panel (a)
shows the radial flux captured early in the evolution of the active
region (AR) where the configuration is simple, with one positive and
one negative footpoint corresponding to the ends of a single flux
loop. Later, as shown in Fig.~\ref{fig:magnetograms}(b), the AR
develops a complex pattern of positive and negative spots that exhibit
shearing.  This configuration indicates that the field lines in the
chromosphere have become twisted, and this may lead to a build-up of
magnetic energy, to magnetic reconnection, and to a solar eruption.
Indeed, a high-intensity flare was generated in this active region 24
hours after the observation in panel (b).  Panel (c) shows the radial
post-flare flux, which is still complicated though perhaps less
sheared than panel (b).  This reduction in complexity is expected
after the release of magnetic energy from the eruption.

For the study reported here, we choose SHARPs records from May 2010 to
December 2017, using a one-hour cadence to reduce redundancy between
consecutive images.  This data set contains a total of $505,872$
records.  For training the ML models, each image is labeled based on
whether the corresponding AR produced a flare in the next 24 hours.
This information is not part of the HMI data set, so we use the NOAA
Geostationary Operational Environment Satellite (GOES) X-ray
Spectrometer flare catalog,\footnote{\scriptsize \tt
www.ngdc.noaa.gov/stp/space-weather/solar-data/solar-features/solar-flares/x-rays/goes/xrs/}
which contains records of the location, time, and magnitude of all
solar flares since 1975.  This catalog categorizes flares into five
classes based on their intensity, from the weaker A, B and C,
to the stronger M and X class flares.
For our period, the GOES database reports 509 M-class flares and 36
X-class flares.  Focusing on the stronger flares because of their
potentially catastrophic consequences, we label each SHARP as flaring
if an M1.0+ flare (one with an intensity above $10^{-5}
\frac{W}{m^2}$) occurred from it within 24 hours after the observation
time; otherwise, it is labeled as non-flaring.
This yields a data set containing 3872 active regions, with $453,273$
SHARPs labeled as non-flaring and 5769 as flaring.  This kind of
imbalance is a major issue from a machine-learning point of view, as
we will discuss further below.

Training, optimizing and evaluating ML models requires splitting the
data into training, validation and testing sets respectively.  We
assign all images of a given AR to a single set, rather than choosing
the images for each set randomly.  This ensures that observations of
the same AR at different points in its evolution do not appear in both
the training and testing sets, thus preventing a possible artificial
score improvement from testing the model on data related to that used
for training.  Of the 3872 active regions, 70\% were used for training
(for fitting the model to the data), 10\% for the validation (for
selection of optimal hyperparameters), and 20\% for the testing (to
evaluate the trained and tuned models on previously unseen data). We
produce ten different training-validation-testing splits through
randomized selections, each generated with a different seed. These
serve as ten trial runs of the experiment of training, optimizing, and
testing our models.

\section{Featurization of Magnetograms} 
\label{sec:features}

The raw magnetogram data described in the previous section (e.g., the
value of $B_r$ at every pixel in a SHARPs record) can be used
directly to train models like convolutional neural networks, which
were developed for the specific purpose of learning patterns in
images.  Alternatively, the magnetogram data can be preprocessed to
extract meaningful numerical attributes that can then be used to train
other types of ML models, including support vector machines, extremely
randomized trees, or logistic regression models.  Feature
engineering---the task of crafting a set of attributes that help the
ML methods learn better---can be a real challenge.  The more salient
the features are in the context of the task at hand, the more traction
they give the method, but it is not always obvious which attributes to
choose.  Domain knowledge can be useful in this endeavor, but
unexpected attributes can also be predictive, particularly in the
context of complicated problems and rich data.

Here, we work with two different feature sets: (i) a set of
standardized physics-based properties used by many solar flare
prediction methods and (ii) a set of abstract properties derived using
an shape-based featurization method first proposed in
\citep{Deshmukh2020}. In addition we also consider a third feature
set that combines (i) and (ii).

\subsection{Physics-based features}
\label{sec:sharps_features} 

Each record in the SHARPs data set includes a set of metadata
containing values for a variety of numerical attributes that represent
properties of the corresponding active region.  These attributes,
developed and refined by the solar physics community over the past 11
years, are given in Table~\ref{tab:sharps_features}.
\begin{table}
	\begin{center}
		\begin{tabular}{| l | l | l |}
			\hline
			Acronym & Description & Units\\
			\hline
			LAT\_FWT & Latitude of the flux-weighted center of active pixels & $degrees$ \\
			LON\_FWT & Longitude of the flux-weighted center of active pixels & $degrees$ \\
			AREA\_ACR & Line-of-sight field active pixel area & $micro$ $hemispheres$ \\
			USFLUX &  Total unsigned flux & $Mx$ \\
			MEANGAM &  Mean inclination angle, gamma & $degrees$ \\
			MEANGBT &  Mean value of the total field gradient & $G/Mm$ \\
			MEANGBZ &  Mean value of the vertical field gradient & $G/Mm$ \\
			MEANGBH &  Mean value of the horizontal field gradient & $G/Mm$ \\
			MEANJZD &  Mean vertical current density & $mA/m^2$ \\
			TOTUSJZ &  Total unsigned vertical current & $A$ \\
			MEANALP &  Total twist parameter, alpha & $1/Mm$ \\
			MEANJZH &  Mean current helicity & $G^2/m$ \\
			TOTUSJH &  Total unsigned current helicity & $G^2/m$ \\
			ABSNJZH &  Absolute value of the net current helicity&  $G^2/m$ \\
			SAVNCPP &  Sum of the absolute value of the net currents per polarity & $A$ \\
			MEANPOT &  Mean photospheric excess magnetic energy density & $ergs/cm^3$ \\
			TOTPOT &  Total photospheric magnetic energy density & $ergs/cm^3$ \\
			MEANSHR &  Mean shear angle (measured using $B_{total}$) & $degrees$ \\
			SHRGT45 &  Percentage of pixels with a mean shear angle greater than 45 degrees & $percent$ \\
			R\_VALUE & Sum of flux near polarity inversion line & $G$ \\
			\hline
		\end{tabular}
	\end{center}
	\caption{The SHARPs feature set. Values for these 20 features, together
          with error estimates, are available for each magnetogram in
          the SDO HMI database.  Abbreviations: $A$ and $mA$ are
          Amperes and milli-Amperes, respectively; $Mm$ is megameters,
          $G$ is Gauss and $Mx$ is Maxwells.}
	\label{tab:sharps_features}
\end{table}
They are predominantly derived from the raw magnetic flux
observations and are believed to be useful indicators of solar flares.
Their values are calculated automatically and stored by the Joint
Space Operations Center group at
Stanford.\footnote
{http://jsoc.stanford.edu/}
The vast majority of ML-based flare forecasting work has, as mentioned above, used these 20
quantities as the feature set.

\subsection{Shape-based features}
\label{sec:tda}

The evolution of the magnetic fields in the Sun during the lead-up to
a solar flare manifests as an increase in the complexity of the
structures on the magnetogram.  This observation, which is visually
obvious from the shapes of the regions in
Fig.~\ref{fig:magnetograms}, plays a critical role in the qualitative
classifications used in operational space-weather forecasts.  The
McIntosh classification system used at the NOAA Space Weather
Prediction Center, for instance, is based on characteristics like the
presence of umbras and penumbras.
These spatial details are, however, largely absent from the
definitions of the SHARPs features of Table~\ref{tab:sharps_features},
most of which are aggregate quantities such as means or totals.
Thus it seems natural to explore whether features that characterize
shape would be useful in ML-based flare forecasting methods---not only
because this is what human forecasters use in their classifications,
but also because AR shape in the photosphere has fundamental,
meaningful connections to the coronal magnetic field physics leading
up to an eruption.

Topology is the fundamental mathematics of shape: it distinguishes
sets that cannot be deformed into one another by continuous
transformations.  Part of this shape
classification---homology---corresponds to the number of connected
components, 2D holes, 3D voids, etc., of a set. These numbers are
known as the Betti numbers, $\beta_0, \beta_1, \beta_2, \ldots$, where
$\beta_k$ is the number of $k$-dimensional ``holes.''  Topological
data analysis (TDA), or computational topology, operationalizes this
framework for situations where one has only finitely many samples of
an object.  One way to create a shape from finite data is to ``fill in
the gaps'' between the samples by treating two points as connected if
they lie within some distance $\epsilon$ of one other.  Building an
approximating object in this fashion, 
TDA computes the Betti numbers, then varies $\epsilon$ and repeats the
process.  The dependence of $\beta_k$ on $\epsilon$ provides a rich
morphological signature that captures the shape of an object at multiple
resolutions.  One can construct an even-richer representation by tracking
the $\epsilon$ value at which each feature is formed, and at which it
is destroyed or merges with another.  This results in a set of
$(birth, death)$ values of $\epsilon$ for each feature (component, 2D
hole, 3D void, etc.).
This methodology has proved to be quite powerful; it has been
successfully applied to a range of different problems ranging from
coverage of sensor networks \citep{deSilva07}, to structures in
natural images \citep{Ghrist08}, neural spike train data
\citep{Singh07}, and even the large-scale structure of the universe
\citep{Xu2018}.  

Our strategy for employing TDA in solar-flare forecasting, originally
proposed in \cite{Deshmukh2020}, is somewhat different from the
approach described in the previous paragraph.  Conjecturing that the
shapes of the level sets of a magnetogram are of central importance
in this problem,
we use the magnetic field intensity $B_r$ as the variable parameter,
rather than a distance $\epsilon$.  We threshold the SHARPs image,
keeping only the pixels where the magnetic field intensity falls at or
below some value (i.e., sub-level thresholding) then compute the
topology of the resulting object.  By varying the threshold and
tracking the birth and death of each feature, we obtain a signature
that captures the morphological richness of a magnetogram in a manner
that factors in the field strength as well as its spatial structure.
A brief synopsis of this TDA-based feature extraction process follows;
more details are given in \cite{Deshmukh2020}.

We build what is technically known as a cubical complex
\citep{Kaczynski2004} from the pixels in the SHARPs image whose $B_r$
values fall below some threshold.  Pixels are connected in such a
complex if they share an edge or a vertex.  Since magnetograms are 2D
images, only connected components and 2D holes (i.e., non-contractable
loops) make sense---there is no higher-dimensional structure.
Counting these gives $\beta_0$ and $\beta_1$ for the given threshold
field, and this computation is then repeated for a range of $B_r$
thresholds.

\begin{figure}
	\centering
		\includegraphics[width=0.90\textwidth]{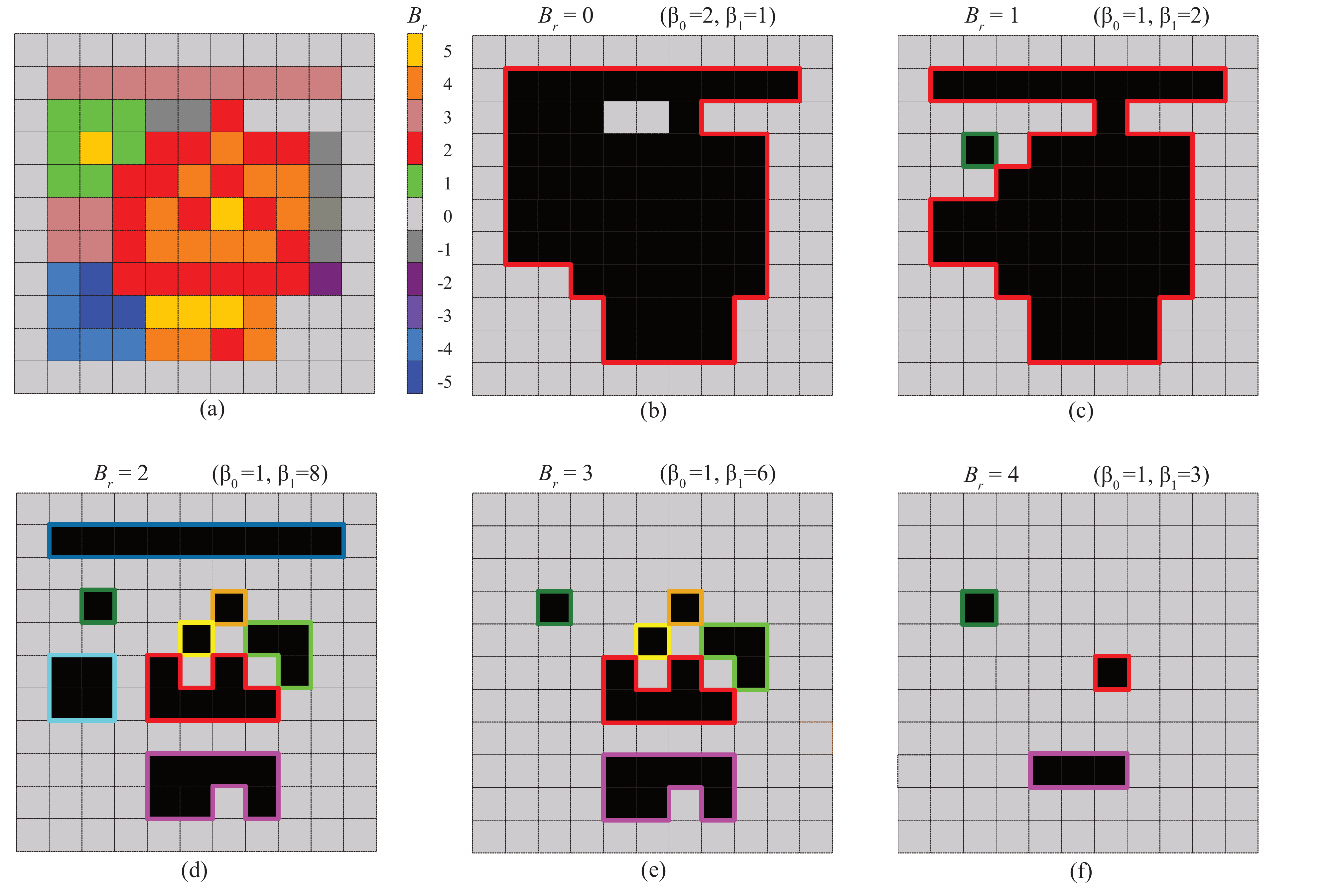}
\\
	\caption{Topological data analysis for the
		example ``image'' in panel (a). The gray pixels in panels (b)-(f) represent the
		cubical complex of the image for five sub-level
		thresholding values, $B_r= 0,1,2,3$, and $4$.  For each complex,
		the $(\beta_0,\beta_1)$ values are given. The colored loops represent
		the holes in the thresholded images.
}
	\label{fig:components} 
\end{figure}
Figure~\ref{fig:components} demonstrates this procedure for a simple
example. Panel (a) represents an image, with each pixel color-coded
according to intensity. Given an intensity value, a sub-level
thresholded image corresponds to those pixels with a magnitude at or
below the threshold. The gray regions in in panels (b)-(f) represent
such images for thresholds $B_r \in [0,4]$. Pixels that share an edge
or a vertex in a thresholded image become a component, which
contributes to $\beta_0$.  Empty regions (black) in the interior of a
thresholded image that are surrounded a loop of connected gray pixels
become holes, incrementing $\beta_1$.  In panel (b), where $B_r = 0$,
the gray, thresholded image contains two components separated by the
empty, black region where the intensity is larger than zero; thus
$\beta_0 = 2$.  There is a single non-contractable loop in the image
(the red curve) that encloses the empty region, so $\beta_1 = 1$.
Increasing the threshold to $B_r = 1$, as in panel (c), causes the
image to enlarge, shrinking the hole and splitting it into two; thus
$\beta_1 = 2$.\footnote{The green box corresponds to a loop in the
image since gray pixels are connected at vertices.}  The two
components from panel (b) merge in panel (c), and for the remainder of
the thresholding process, the number of components remains one, so
$\beta_0 = 1$.  Upon raising the threshold, as shown in panel (d), the
dominant hole splits into seven while the single-pixel hole remains
intact, bringing the number of holes to eight ($\beta_1 = 8$). At $B_r
= 3$ in panel (e), two holes disappear or \textit{``die''}; thus
$\beta_1$ = 6. Finally, for $B_r$ = 4 in panel (f), the number of
holes is reduced to three as three of the holes are filled in by new
image pixels.  If the threshold were raised to five (not shown), all
pixels would be filled so that the image would have $\beta_0 = 1$ and
$\beta_1 = 0$.

\begin{figure}
	\centering
		\includegraphics[width=0.40\textwidth]{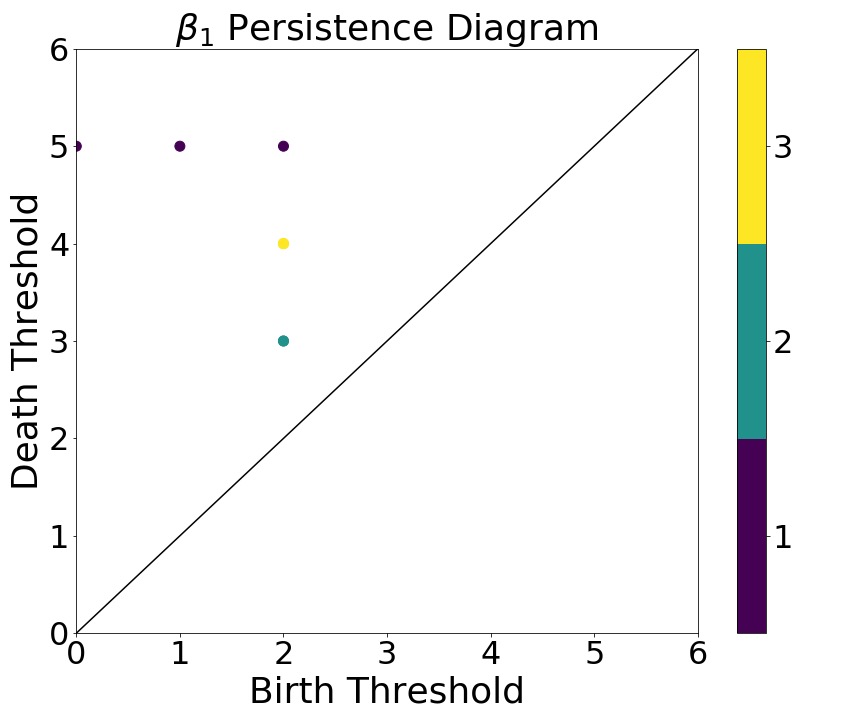}
	\caption{The $\beta_1$ persistence diagram for
          Fig.~\ref{fig:components}.  The color scale represents the
          multiplicity, i.e., how many points share the same $(birth,
          death)$ values.}
	\label{fig:PDExample} 
\end{figure}

A number of different representations have been developed by the TDA
community to capture the information about the scale and complexity of
the different structures that is produced by such an analysis
\citep{Ghrist08}. One of the most common, the persistence diagram (PD)
\citep{Edelsbrunner2000}, is a plot of the birth and death
values for each feature (e.g., the threshold value of $B_r$).  Points
that lie far from the diagonal on such a diagram are said to be
\textit{persistent}, as their birth and death values are widely
separated.  Points near the diagonal, which have short lifespans, are
often formed due to noise in the data \citep{Ghrist08}.
Figure~\ref{fig:PDExample} shows a $\beta_1$ PD for the example of
Fig.~\ref{fig:components}.
Each point in this PD corresponds to the $(birth, death)$ thresholds
for a hole in the filtered image.  We use color to indicate the number
of holes with same lifespan.  In our example, there are three holes
that live until $B_r = 5$.  The longest-lived hole, indicated by the
red border in Fig.~\ref{fig:components}, is born at $B_r = 0$, and
thus corresponds to the point (0,5) on the PD.  The single-pixel hole
(green loop) is born at $B_r=1$, giving the point $(1,5)$ on the PD;
the hole with the magenta border is born at $B_r = 2$ so it appears on
the PD at $(2,5)$.  There are also five relatively short-lived holes;
three of them (with yellow, orange and light-green borders) are
represented by the yellow point $(2,4)$ on the PD.  The remaining two
are the shortest-lived (cyan and blue) and correspond to $(2,3)$.

One can similarly construct a $\beta_0$ persistence diagram to capture
the birth and death of the connected components in the analysis.  For
the example here, this is not too interesting, since apart from the
brief appearance of a second component at $B_r = 0$, there is only one
component for all thresholds.  This holds true for the $\beta_0$ PDs
of actual magnetograms as well---they do not add much information to
the analysis.  For this reason, we restrict further analysis to
$\beta_1$ PDs.

\begin{figure}
\begin{center}
\includegraphics[height=0.32\textwidth]{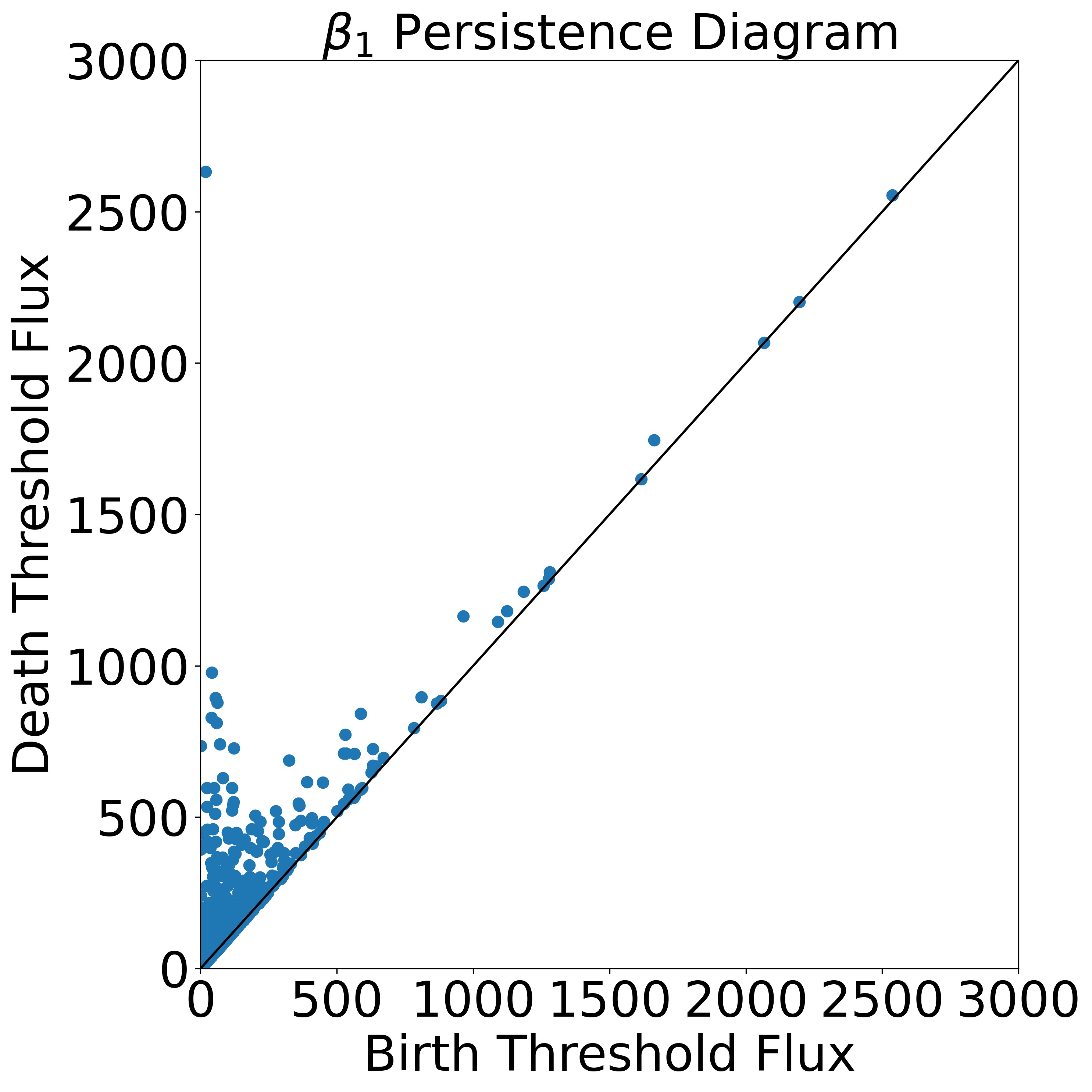}
\includegraphics[height=0.32\textwidth]{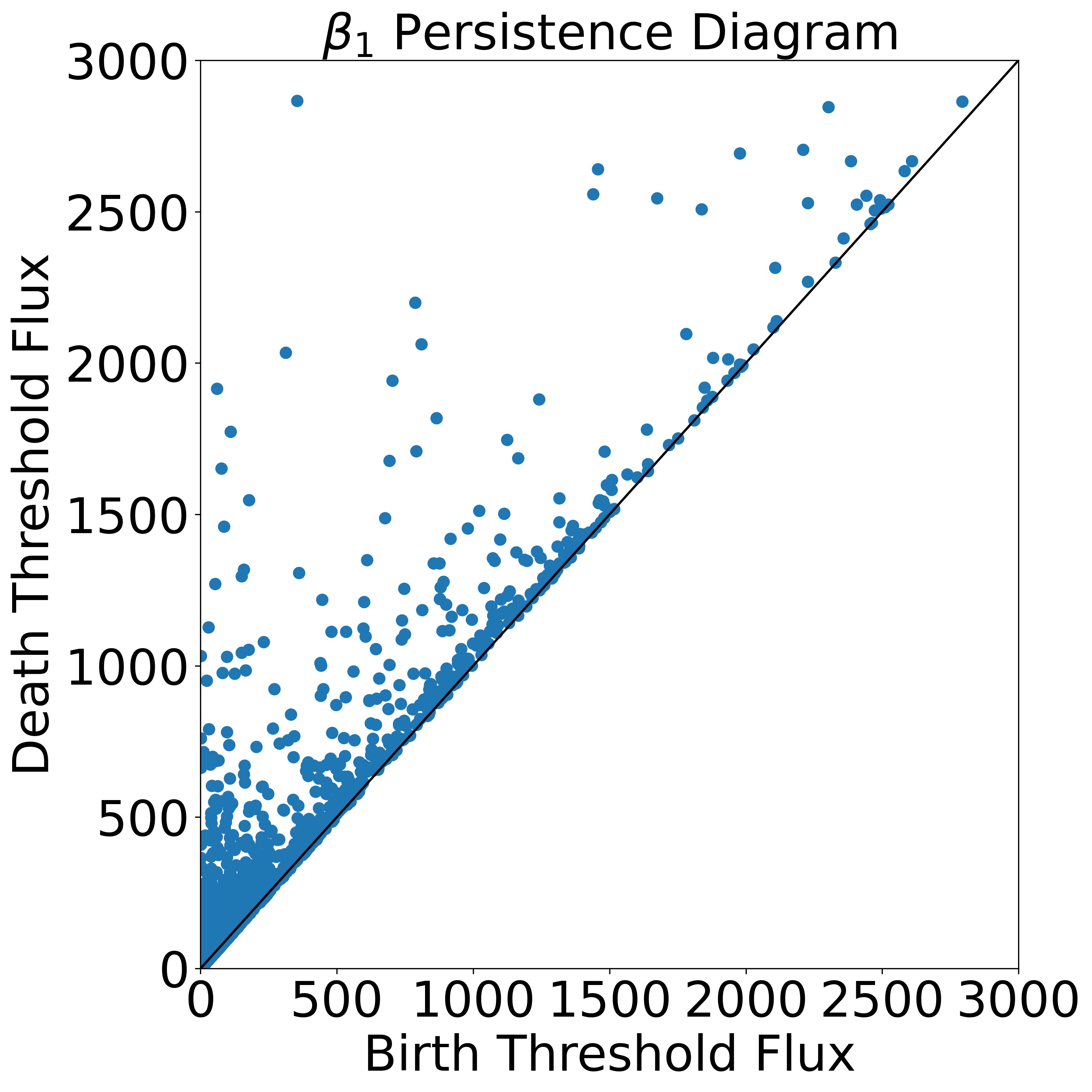}
\includegraphics[height=0.32\textwidth]{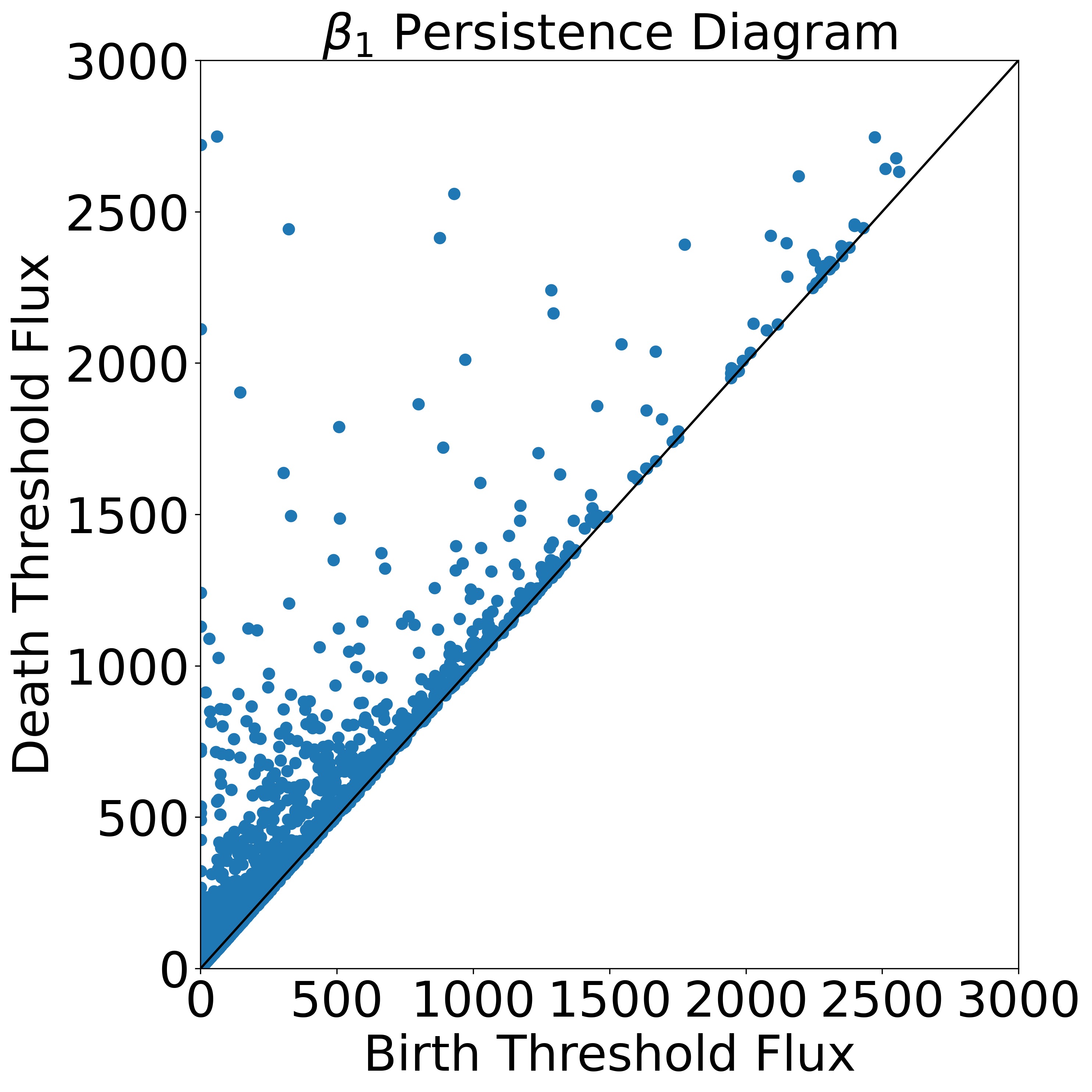}	
\end{center}
\caption{$\beta_1$ persistence diagrams generated from the positive
  magnetic flux density values of the magnetograms of
  Fig.~\ref{fig:magnetograms}.  A clear change in the topology of
  the field structure is observed well in advance of the major flare
  eruption that occurred in this AR at 0910 UT on 6 September 2017.}
\label{fig:pds}
\end{figure}

Persistence diagrams of real-world images can be far more complicated.
Figure~\ref{fig:pds} shows $\beta_1$ PDs for the AR from
Fig.~\ref{fig:magnetograms}.
As in Fig.~\ref{fig:components}, these persistence diagrams show only
the positive sub-level thresholds for $B_r$. The three images in
Fig.~\ref{fig:pds} clearly bring out the evolving structure of this
active region.  The relatively simple structure of the fields
in Fig.~\ref{fig:magnetograms}(a) creates one long-lived hole in
Fig.~\ref{fig:pds}(a) at approximately (0, 2600), corresponding to the
central white spot, along with a large number of short-lived holes
near the diagonal, which correspond to short-lived structures likely
due to noise.  The persistence diagram in panel (b), 24 hours before
an X-class flare, has a large number of long-lived holes, reflecting
the complexity of the structure of the active region at this time.
After the flare, the PD in the rightmost panel is still complex, but
the number of persistent holes has decreased; for example, the number of holes
with a lifespan of 500 Gauss or more drops from 70 in
panel (b) to 61 in panel (c).

We can also construct separate PDs for
the negative flux regions by choosing negative threshold values,
though in this case it is appropriate to use ``super-level'' instead
of sub-level thresholding. Thus for this case we first include---for
the threshold $B_r =0$---all pixels with $B_r \ge 0$, and then filter
for increasingly negative values of $B_r$. For example, in
Fig.~\ref{fig:components}(a), this process will leave a hole
surrounding the blue pixels for a super-level threshold of $B_r = -3$.
For the AR of Fig.~\ref{fig:magnetograms}, the resulting PDs (not
shown) exhibit a similar evolution pattern to those in
Fig.~\ref{fig:pds}.

These results suggest that persistence diagrams can be useful
indicators of impending solar flares.  To operationalize this in the
context of machine learning, however, there is an additional
challenge.  ML models generally require data that have a fixed
dimension, but the PD contains an arbitrary number of $(birth, death)$
tuples.  Various approaches to this ``vectorization'' problem have
been proposed \citep{Bubenik2015, Adams2017, Chazal2014, Bubenik2015,
  Reininghaus2015, Kusano2016, Carriere2017, Le2018, Carriere2019}.
Here we use a simple technique: we choose $10$ equally spaced
thresholds with $B_r >0$ for sub-level thresholding and $10$
thresholds with $B_r < 0$ for super-level thresholding, setting the
maximum $|B_r| = 5000G$; this gives a range that covers the magnetic
flux observed in most active regions.\footnote
{Specifically the thresholds are $\lbrace 263G, \cdots, 4473G, 5000G \rbrace$
for sub-levels and  $\lbrace -263G, \cdots, -4473G, -5000G \rbrace$
for super-levels.}
The spacing of the thresholds is chosen to ensure a good balance
between redundancy (i.e., closely spaced, highly similar images with
identical $\beta$ counts) and adequate representation (avoiding a
spacing so coarse as to miss important information).  This process
produces two $10 \times 10$ discrete PDs, one for positive and one for
negative $B_r$.  To telescope this information into a form that is
usable by ML methods, we construct a $10$-dimensional vector from each
PD by simply counting the number of ``live'' holes at each threshold.
These 20 values make up the feature set for the results presented in
Section~\ref{sec:Results}.
\subsection{Feature-Set Reduction}
\label{sec:pca}

As part of the training process, ML models learn which attributes of
the input data, in which combinations, are meaningful.  Their
efficiency in doing so depends on the amount and complexity of the
data, and also---importantly---on the way it is represented.  In
general, a larger feature set leads to increased model complexity,
which in turn can result in overfitting and increased computational
time \citep{Goodfellow2016}, as well as requiring larger training data
sets. The nature of the individual features also matters.  Features
that are not salient, or that are redundant, slow down the learning
process.  For these reasons, it is important to minimize the number
of features and maximize their relevance.

A good way to approach this problem is to apply
dimensionality-reduction techniques to the feature space in order to
find the most relevant subspaces.  A variety of methods have been
proposed for this, including principal component analysis, linear
discriminant analysis, t-distributed stochastic neighbor embedding
\citep{Maaten2008}, uniform manifold approximation and projection
\citep{McInnes2018}, etc.  We use principal component analysis (PCA)
because of its simplicity and effectiveness.  It determines an
alternative basis set to represent the data by iteratively
constructing an orthogonal basis such that the variance of the data
along the first dimension is maximal, the second dimension is in the
direction of maximum variance that is orthogonal to the first
dimension, and so on.
One then typically keeps the first $l$, say, principal vectors that
together account for some chosen fraction of the total variance.
This approach, applied to an $n$-dimensional feature set, 
effectively reduces the dimensionality to $l$.
Note that each of the $l$ basis vectors may contain all of the original $n$ values;
we will discuss this more below.

\begin{figure}
\centerline{
	\includegraphics[width=0.5\linewidth]{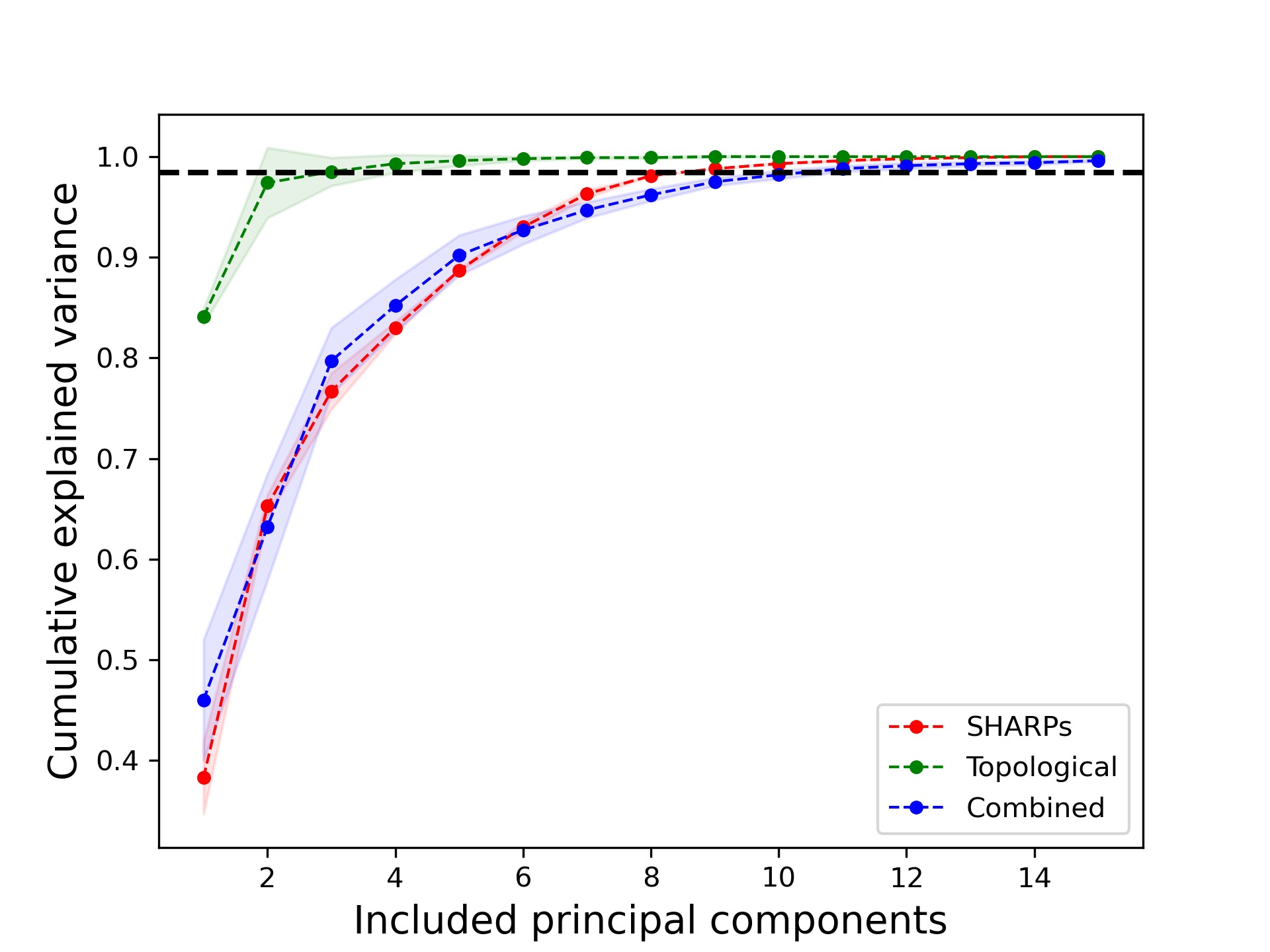}}
	\caption{Cumulative explained variance plots of the principal
          components for the three feature sets, determined from the
          ten training sets.  The darker curves represent the medians
          of the explained variance, while the shaded regions around
          them indicate standard deviations.
          The dashed line marks the 98.5\% level.}
	\label{fig:explained_variance}
\end{figure}

Applying this approach to the ten randomly shuffled training sets
discussed in Sec.~\ref{sec:data}---first using the SHARPs feature set
of Sec.~\ref{sec:sharps_features}, then the shape-based feature set of
Sec.~\ref{sec:tda}, and finally their combination---gives
Fig.~\ref{fig:explained_variance}.
For the SHARPs feature set, the first principal component captured
38\% of the variance.  As Fig.~\ref{fig:explained_variance} shows,
adding a second principal component increases the total to 60\%.  The
topological feature set is much more anisotropic: its first principal
component captures almost 85\% of the variance of the data set, a
level that requires four principal components in the SHARPs feature
set.  This is a direct reflection of the relevance of the TDA features
for the purpose of flare prediction---as well as an indication of how
many principal components will be needed as discriminating features
for the corresponding ML models.  To that end, we apply a threshold of
98.5\% (the dashed line in Fig.~\ref{fig:explained_variance}) to
select nine, three, and eleven dimensions from the orthogonal PCA
representation for the reduced-dimension versions of the traditional,
topological, and combined feature sets, respectively.

\begin{figure}
\centering
\gridline{
	\fig{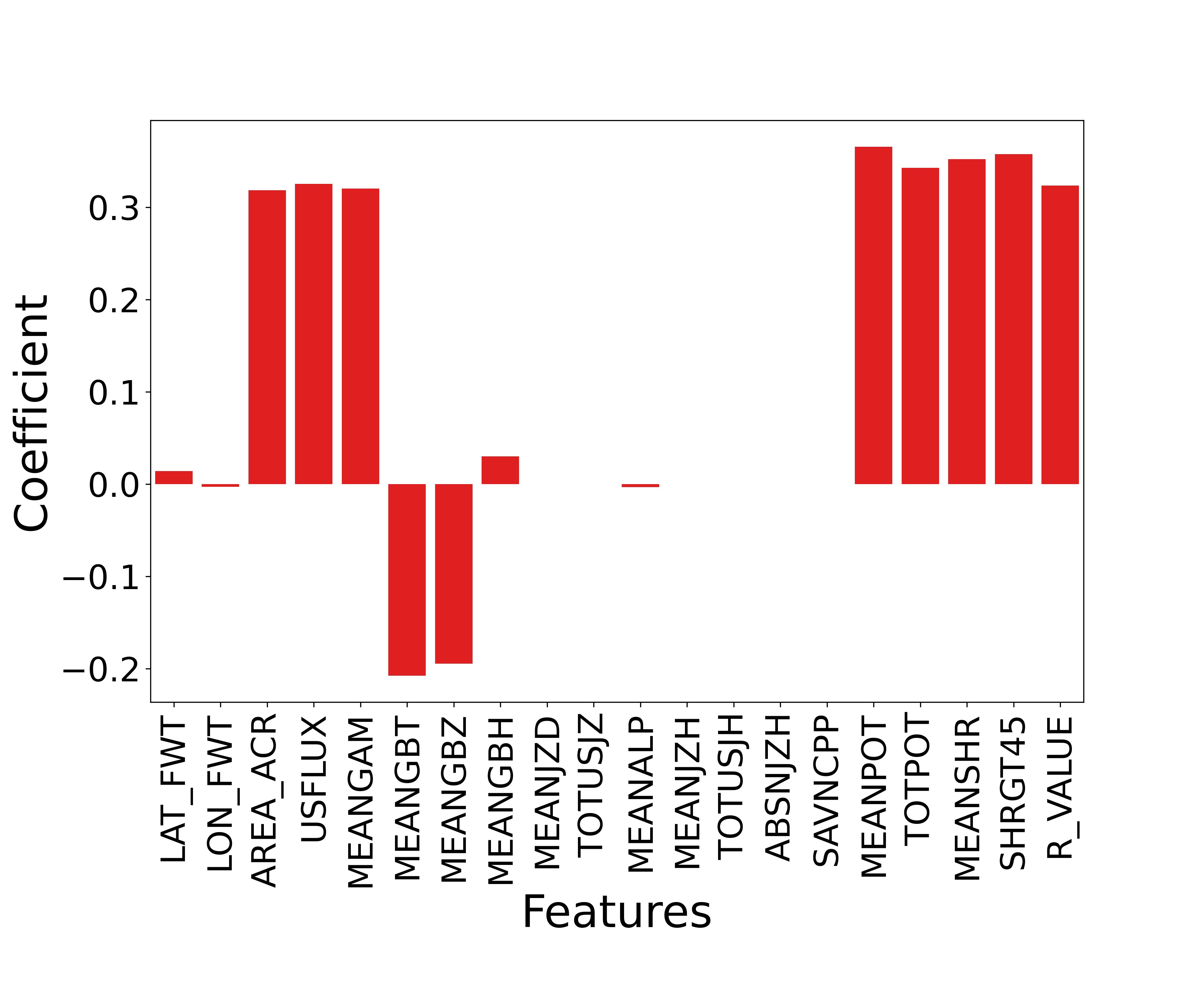}{0.45\textwidth}{(a) SHARPs features set}
	\fig{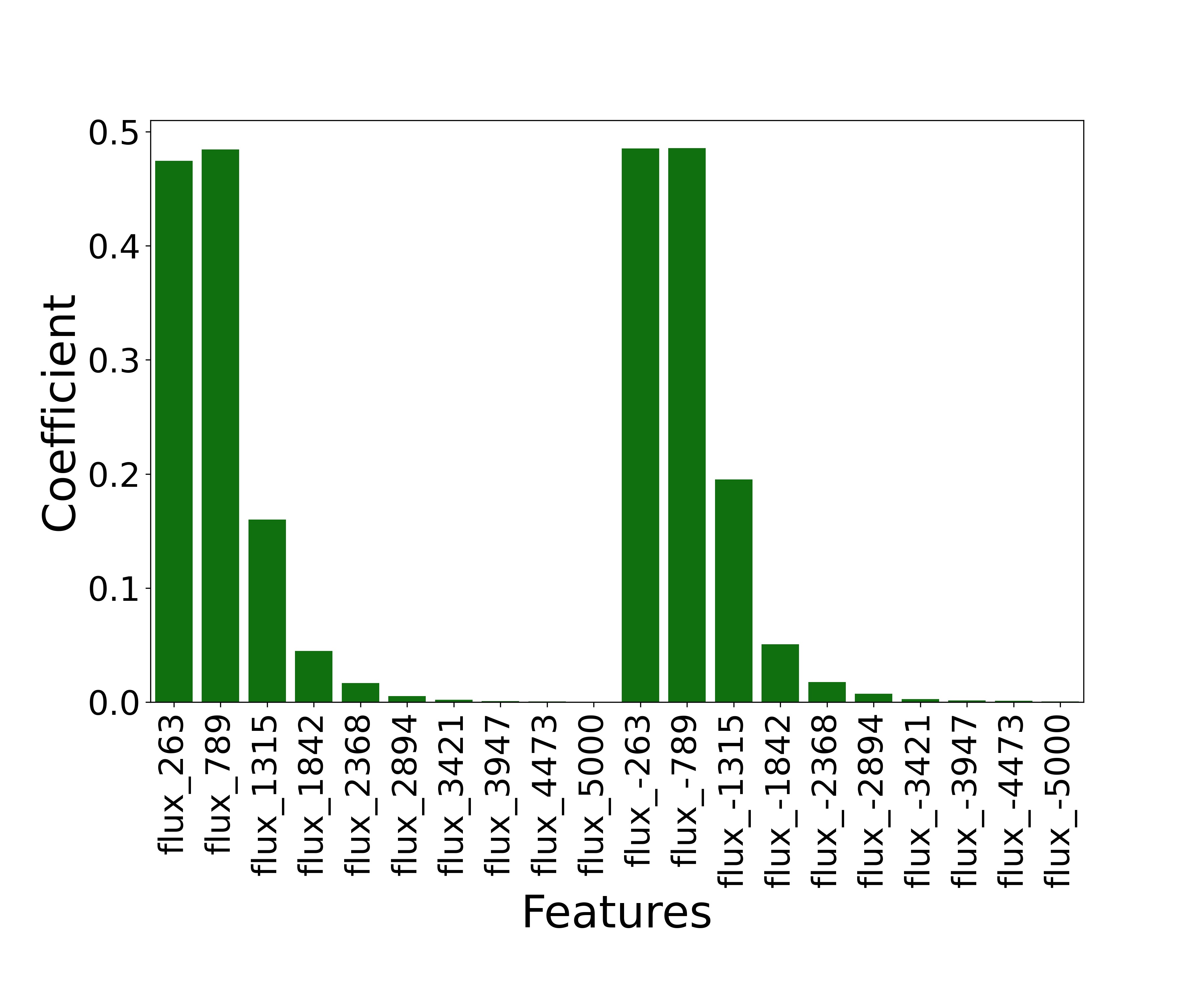}{0.45\textwidth}{(b) Topological feature set}
}
\gridline{
	\fig{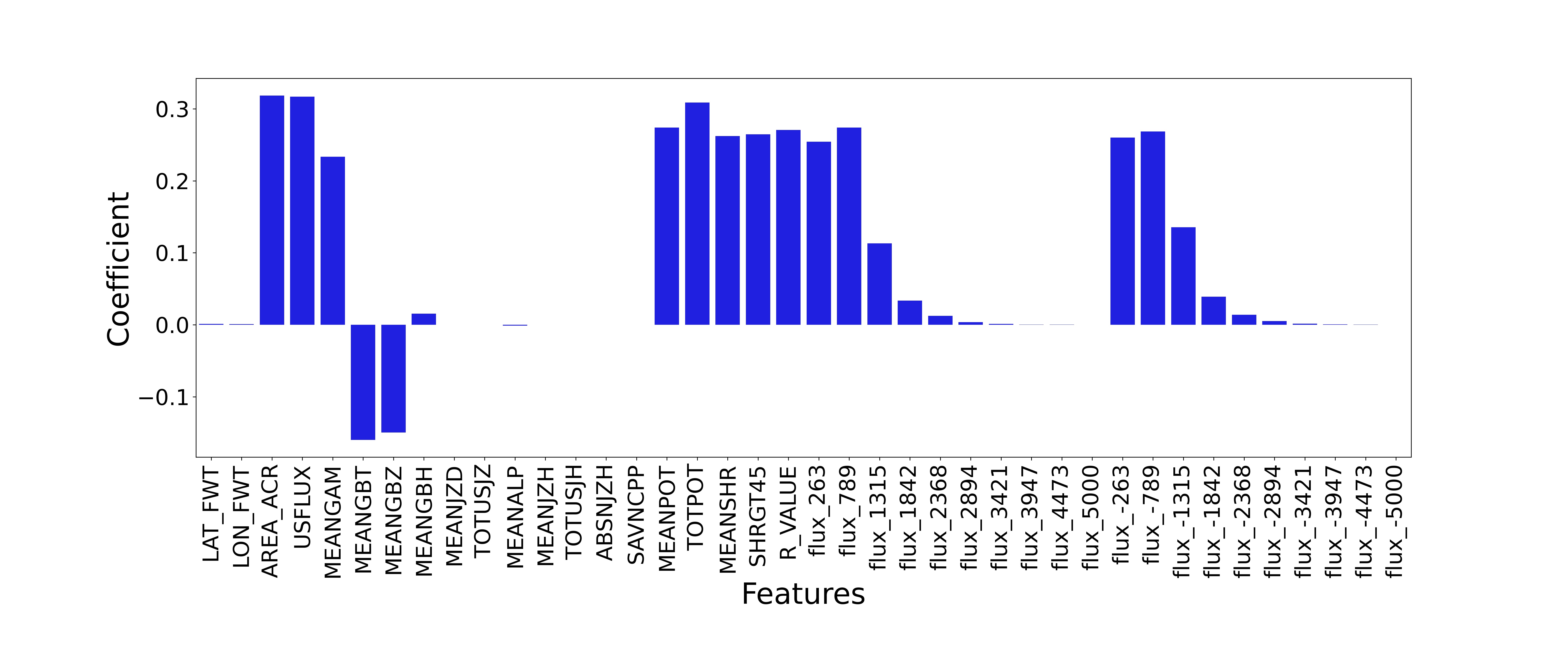}{0.9\textwidth}{(c) Combined feature set}}
\caption{The weights of the SHARPs, topological, and combined features
  in the first principal component of the corresponding feature set
  for one of the ten training sets examined in this paper.  Labels of
  the topological features indicate the flux level of the threshold
  value used to construct those features.
}
\label{fig:PCA-details}
\end{figure}

A detailed analysis of the reduced features is revealing.  The
coefficients of the first principal component---the weights of
each SHARPs attribute in the basis vector---for
one of the ten training sets are shown in
Fig.~\ref{fig:PCA-details}(a).
These results are to some extent consistent with the science: the {\tt
  R\_VALUE} and {\tt USFLUX} attributes, which are known
to be important for flare prediction, are weighted heavily in the
first principal component of the SHARPs set.  Note, though, that six
other quantities are also weighted heavily in
Fig.~\ref{fig:PCA-details}(a); moreover, this first principal
component does not capture much of the variance, so one should not
over-interpret its weights.  Rather, one must think about \textit{all}
of the principal components for each feature set, acting together as a
basis set for the new feature space.  (A feature that is totally
absent from every principal component, of course, is certainly not
salient, but a low-weighted feature in the first one could have a high
weight in another.)  That being said, interpretation of
Fig.~\ref{fig:PCA-details}(b) is somewhat different because, as
Fig.~\ref{fig:explained_variance} shows, the first principal component
in the topological feature set captures a very large fraction of the
variance.  
It is interesting to note that the topological features
at lower magnetic fluxes are weighted more heavily in the first
principal component, and with some $+/-$ symmetry.  This makes sense
in view of the nature of the threshold construction:
a high threshold value removes much of the small-scale structure of
the magnetogram, leaving only the highest-magnitude pixels.  Finally,
note that the first principal component from the combined feature set
is a weighted combination of both SHARPs and TDA features.  As we
shall discuss in Section~\ref{sec:Results}, this can have implications
regarding model performance.

Of course, PCA only finds a new basis set for the feature space; it
does not produce a feature ranking.  The previous paragraph only
explains how one principal component is represented in terms of the
original features, in one particular split of the data.  Even though
this gives possible ties to the physics, it does not represent a
comprehensive, formal analysis of feature importance in the
classification problem.

\section{Machine Learning: Models and Framework} \label{sec:models}

Evaluation and comparison of the predictive power of different
machine-learning methods and different feature sets require care.  The
first step, as described at the end of Section~\ref{sec:data},
involves creating a \textit{split} of the data set into training,
validation, and test sets that are representative of the problem at
hand, but do not artificially boost the performance of the model.
Secondly, it is important to work with a \textit{range} of ML models
in order to be able to make general claims about whether one feature
set is more useful than another.  The models, as described
in Section~\ref{sec:MLModels}, were chosen to span ranges of complexity and
strategies.  Thirdly, ML methods have a number of
\textit{hyperparameters} that guide their training processes.  Optimal
values for these parameters depend on the model and the data, so a
truly fair comparison requires individualized tuning on a case-by-case
basis.  To that end, we use a $k$-fold cross-validation approach to
automatically determine hyperparameters from the data
as explained in Section~\ref{sec:HyperparameterTuning}. This
systematic approach, which is in contrast to the trial-and-error
approach used in the majority of the flare-prediction literature,
ensures that each model achieves peak performance for each feature
set.

\subsection{Machine Learning Models}\label{sec:MLModels}

In our study, we use four different models: logistic regression,
multilayer perceptrons, long short-term memories, and extremely
randomized trees.  
In the following paragraphs, we give brief descriptions of these
methods; for more details, please see \cite{Murphy2012,
Goodfellow2016} or any other basic machine-learning reference.

\smallskip


{\bf Logistic regression}, perhaps the simplest of all models in the
ML literature, uses a sigmoid function
$h: \mathbb{R}^n \to [0,1]$ as the basis function to 
the model the data:
\[ 
	h_\theta(x) = \frac{1}{1 + e^{-\theta^Tx}} .
\]
The vector \(\theta\) corresponds to weights are applied to each
element of the input vector $x$.  In the
context of the flare-prediction problem, $h_\theta(x)$ represents the
flaring probability, which we then convert to a categorical
``flare/no-flare'' output using a threshold of 0.5---if $h_\theta(x) < 0.5$ 
then $x$ is classified as non-flaring, otherwise it is classified as flaring.
Training this model is a
matter of determining values for \(\theta\); we accomplish this using
the LBFGS algorithm of \citep{Liu1989}.  The only hyperparameter
involved in this process is the class weight that is used in the
gradient descent: the higher the weight for one class, the more the
model is penalized for getting the classification wrong.


A {\bf multilayer perceptron} or MLP is a type of feedforward
artificial neural network containing multiple layers of nodes or
\textit{neurons}, with the outputs of each layer propagated forward to
the next layer.  These canonical ML models contain an input layer,
some number of hidden layers, and an output layer.  The output of each
neuron is a nonlinear function of its inputs with a single free
parameter, typically a multiplicative constant, that is ``learned''
during training using some optimization strategy (e.g., gradient
descent) on a suitable loss function: in this case, the weighted
binary cross-entropy loss function, as described in
\cite{Deshmukh2020}.  We employ the Adagrad optimizer
\citep{Duchi2011} to update the weights during gradient descent and an
architecture with five dense layers containing 36, 24, 16, 8 and 2
nodes, respectively. This configuration was chosen using a manual
trial-and-error approach.  There are two important hyperparameters
here: the class weight, which plays a similar role as in the logistic
regression model, and the L2 regularization constant in the loss
function, which ensures that none of the neuron weights becomes too
large, which would lead to overfitting.

Active regions evolve over time in ways that are meaningful from the
standpoint of solar physics.  Logistic regression and MLPs cannot
leverage the information that is implicit in a sequence of
magnetograms, as their forecasts are based on individual snapshots.
For that reason, a thorough evaluation of ML-based solar flare
forecasting should include methods that factor in the history of the
observations.  To that end, we use {\bf long short-term memories}
(LSTMs), which are designed to work with temporal sequences of data,
using a feedback loop in a hidden layer that takes the state
calculated in time step \(t_{n-1}\) and feeds it to the same network
for the sample at time step \(t_n\).  The LSTMs used in our study
contain the same number of hidden layers as the MLP discussed above,
but with such a feedback loop incorporated in one of the hidden
layers.  This strategy for propagating information forward in time is
powerful, but it can complicate the training of these models.  Simple
gradient descent, for instance, can be problematic if there are
long-term temporal dependencies in the data, since the gradient can
decay as it is propagated through the time steps.  To address this,
LSTM nodes often incorporate mechanisms called ``forget gates'' that
limit the number of steps through which information is propagated
forward in time.  This limit is an important choice, and one that is
generally tuned by hand for a given model and data set.  We use this
approach and find that a sequence length of ten works well in our
application.  Like MLPs, LSTMs have two hyperparameters that
significantly impact their performance: the weights in the binary
cross-entropy loss function and the L2 regularization constant.

The fourth type of model used in our work is a type of Decision Tree,
a class of ML models that have a tree-like structure, where each
branch represents a decision based on some attribute of the data and
the leaves correspond to the salient classes for the problem at hand
(flaring and non-flaring, in our case).  The input data dictate the
path taken through the tree during the classification process,
eventually routing the outcome to one of these classes.  A major
advantage of this strategy is that its results are an indication of
how well each individual feature is able to divide the data set: a
major step towards explainability, a critical challenge in modern AI.
The main disadvantage of decision trees is their tendency for
overfitting.  One can mitigate this by building an ensemble of
trees using a randomly chosen subset of features 
at each branch point---a so-called ``Random Forest''---and use the
mean or mode of their predictions to classify a sample.  The {\bf
  extremely randomized tree (ERT)} \citep{Geurts2006} that we 
use in this paper is a variant of this approach.
These models have three hyperparameters: the class weight; 
the minimum impurity decrease, which controls when nodes will split; and 
the number of the trees in the ensemble, which mitigates overfitting.

This set of choices covers the full gamut of ML methods, in terms of
architecture (e.g., tree-based or not) and complexity (number of free
parameters),
making it a sensible evaluation set.  

\subsection{Hyperparameter Tuning}
\label{sec:HyperparameterTuning}
Since the performance of an ML model
depends on both the data and the training process, and because that
training process is governed by the model's hyperparameters, a fair
comparison of two different ML models requires careful tuning of their
hyperparameters. To do this,  one evaluates its
performance on a subset of the data, called the validation set, for
different hyperparameter combinations.  The best-performing
combination is then used to train the model on the training set before
it is evaluated on the corresponding test set.  Using distinct subsets
of the data for these three purposes ensures that the processes are
completely independent.  

Instead of
using a single validation set, a better strategy is to perform what is
called a $k$-fold cross-validation.  In this approach, the training
set is divided randomly into $k$ subsets, or ``folds," of roughly
equal size.  For each evaluation of hyperparameters, one
of the $k$ folds acts as the validation set and the remaining $k-1$
folds are merged to become the training set.
After training, the model is then tested on the
1-fold validation set.  This process is repeated using each of the $k$
folds, individually, and the success of the hyperparameter combination
is judged using the mean of the model's performance metric across these
runs.

\begin{figure}[htbp]
	\centering
	\includegraphics[width=0.8\linewidth]{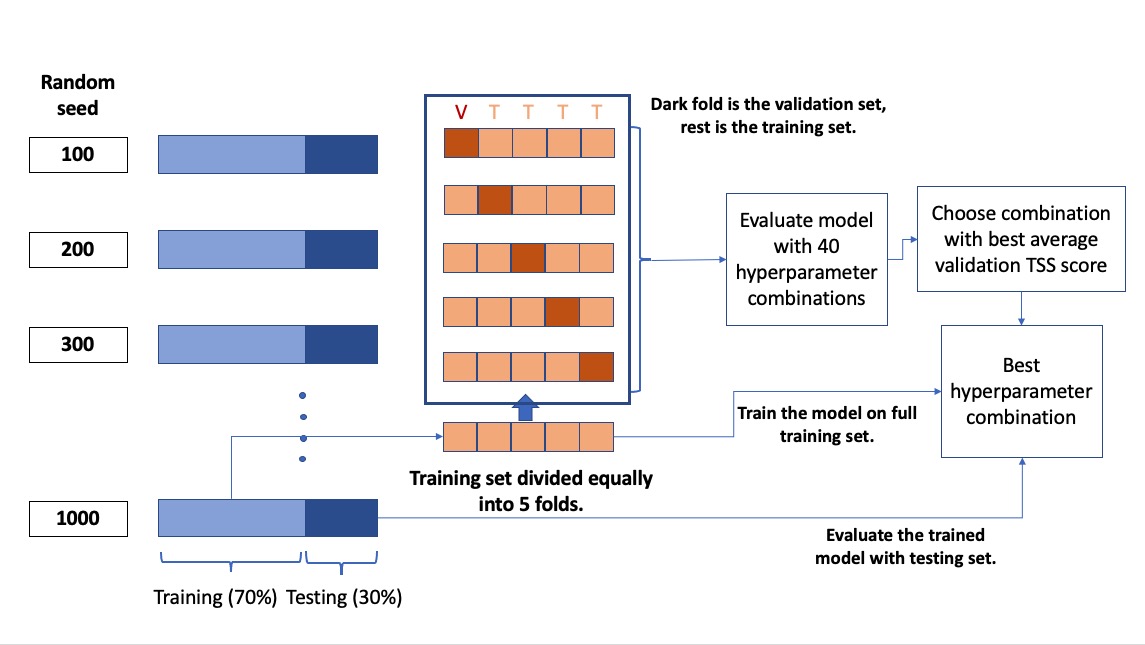}
	\caption{Hyperparameter tuning workflow.}
	\label{fig:hyperparameter}
\end{figure}

This process is illustrated in Fig.~\ref{fig:hyperparameter}.
The first step is the standard random splitting of the data set, as
described at the end of Section~\ref{sec:data}.  We repeat this ten
times using a 70\%-30\% ratio to generate ten trial runs, then
generate $k$ folds on each of those training sets.  In
machine-learning practice, $k$ is generally chosen as $3, 5$, or $10$.
In our case, $k = 5$ works well to produce sufficiently long training
and validation sets.  A second important decision in this tuning
process is the sampling method for the hyperparameter value
combinations.  Various strategies have been proposed for this in the
ML literature: grid search, random sampling, Bayesian sampling, etc.
Here, we use the Bayesian optimization method of \citep{Ruben2014},
that uses Gaussian processes to mix exploration and exploitation in
optimizing the hyperparameter combinations.  We employ the Python
implementation {\tt BayesOptSearch} and use the {\tt ray.tune} Python
library \citep{liaw2018tune} to deploy the sampling and evaluation process
on the graphics processing unit.  As shown in
Fig.~\ref{fig:hyperparameter}, the {\tt BayesOptSearch} method
iteratively samples 40 hyperparameter combinations, choosing the one
that gives the best performance on the validation set, as measured by
some metric that compares the forecast to the ground truth.  (We use
the true skill statistic, as described further in
Section~\ref{sec:Results}.)  For {\tt BayesOptSearch}, we use the
upper confidence bound as the acquisition function, setting the
exploration parameter $\kappa = 5$ in order to balance exploitation
and exploration in the sampling process; see \cite{Snoek2012} for
details.  That sampling process begins with $10$ uniformly chosen
hyperparameter samples, continuing in steps of $10$ to explore $40$
hyperparameter combinations.\footnote
{i.e., using {\tt initial\_random\_steps} in {\tt ray.tune}, with {\tt max\_concurrent\_trials}=10}
The model is then trained on the full training set using the best-performing
hyperparameter values before being evaluated on the 30\% test set.

Not all hyperparameters require this kind of complex, computationally
intensive treatment; some of them can be quite effectively tuned using
manual trial-and-error.  We employ a two-phase model-tuning
strategy, first performing a hand optimization of hyperparameters such as
the number of model layers, the nonlinear activation function, and the
optimization function. These choices are made from among a
finite set of options, and generally impact all metrics in a
similar way.
This is followed by the automated approach described above for tuning
hyperparameters such as the loss function weights or the
regularization penalty.  Automatic tuning of these parameters is
important for two reasons.  Firstly, they take on continuous values,
so hand-tuning becomes cumbersome.  Secondly,
their effects are not independent: changing one often improves
one metric at the cost of another---a situation where an automated,
systematic exploration of the search space can be especially
appropriate.  It is entirely possible, of course, to use the automated
approach for {\sl all} the hyperparameters, but that significantly
increases the computation time.

Hyperparameter tuning is, as should be clear from the details above, a complex process.
We offer these details here not
only so that others can not only fully reproduce these results, but
also use this framework in other applications that involve evaluation
and comparison of flare-forecasting approaches.  It is also worth
mentioning that the metric that one uses to evaluate performance will
affect the process, and in subtle ways: optimizing for pure accuracy,
will produce different results than balancing false
positives against false negatives.  This matter is discussed further
below.

\section{Results} \label{sec:Results}

In this section, we compare the relative prediction performance of the
machine-learning based flare-forecasting models described in
Section~\ref{sec:models} for the three feature sets covered in
Section~\ref{sec:features}.  After extracting values for those
features from the labeled data set described in
Section~\ref{sec:data}, we carry out the hyperparameter tuning
procedure outlined in the previous section on each model/feature-set
combination, using the $k$-fold cross validation approach on the ten
datasets, then train it with optimized hyperparameters on the
corresponding training set.  To evaluate the results, we run the model
on the corresponding test set and compare its 24-hour forecasts to the
ground truth using four standard prediction metrics: accuracy; the
true skill statistic or $TSS$ (also known as the $H\&KSS$),
the Heidke skill score ($HSS_{2}$),
frequency bias ($Bias$),
and $F_1$, which is the harmonic
mean of precision and recall.
These metrics, whose detailed formulae can be found in
\cite{barnes_survey,bobra,Crown:2012}, are derived from the entries of
the contingency table---i.e., the numbers of true and false positives
and true and false negatives.  In the context of this problem, a
flaring magnetogram (or more accurately, a magnetogram with an associated 
M1.0 or larger flare in the next 24 hours) is considered as a positive while a 
non-flaring magnetogram is considered a negative. Since our data set includes
5769 of the former and 447504 of the latter, accuracy is not a very
useful metric here; a simple model that classified every input as
non-flaring would be 98.7\% accurate.  The skill scores
strike various balances between correctly forecasting the positive and
negative samples.  $TSS$ ranges from $[-1, 1]$ and $HSS_{2}$ from
$(-\infty,1]$.  In both cases, these scores are $1$ when there are no
  false positives or false negatives, while a score of $0$ means the
  model is doing only as well as a random forecast, essentially, an
  ``always no-flare'' forecast.  $Bias$ has the range $[0,\infty]$,
  where $Bias<1$ indicates under-forecasting (many false negatives),
  and $Bias>1$ implies over-forecasting (many false positives).  $F_1$
  has the range $[0,1]$ with 1 indicating a perfect forecast score.
  These metrics, all of which are used broadly in the flare-prediction
  literature, span the various methods used to quantify the
  performance of ML models.

As discussed in Section~\ref{sec:models}, hyperparameters for each
model must be individually tuned in order to provide a fair comparison
of their performance.  The choice of metric plays a subtle role here,
since hyperparameters can have different effects on the various
metrics.  Tuning performance based on values of one of them, then, can
impact performance as measured by the others.  The choice of metric is
often left as a decision for the forecaster: some might wish to
prioritize the $TSS$ score, e.g. \citep{Deshmukh2020_IAAI}, while
others might prefer a forecast that has lower false positive rate
\citep{Deshmukh2021Arxiv} or is more reliable \citep{Nishizuka2021}.
In the problem treated here, where the data set is highly imbalanced,
an optimization based on accuracy would be a particularly bad choice,
as it would lead to models defaulting to the ``always no-flare''
forecast.  We use $TSS$ in our work, choosing hyperparameters that
maximize its values via the $k$-fold cross-validation described above.
$TSS$ is a common choice in the flare-forecasting literature, as well
as the broader machine-learning literature.  It does come with
limitations, however: optimizing the TSS score can lead to high false
positives,
thereby impacting some of the other metrics like precision, $F_1$, and
$Bias$.  This effect manifests in the results described below.
 
These experiments were carried out on an NVIDIA Titan RTX (24 GB, 33
MHz) GPU for the deep-learning models (MLP and LSTM),
and on an Intel i9-9280X (3.30 GHz) CPU for the simpler models
(logistic regression and ERT). \footnote{Different machine-learning models
lend themselves to different types of hardware, depending on how well
they parallelize.  The machine used to carry out these experiments
affects only the run time, not the results.}  Run times ranged from
2.5 seconds to train and 0.02 seconds to test each logistic regression
model on the SHARPs feature set to 210 seconds and 9 seconds for the
LSTM model using the combined feature set.
The run time of the hyperparameter tuning procedure ranged from 40 seconds for each logistic
regression model with the SHARPs feature set to just under two hours
for each LSTM model with the combined feature set.

Table~\ref{tbl:models_features} compares the performance of the
various models for the three feature sets: the traditional
physics-based features that appear in the SHARPs metadata
(Section~\ref{sec:sharps_features}), the shape-based attributes
extracted from each magnetogram image using topological data analysis
(Section~\ref{sec:tda}), and a third set that combines the two.
\begin{table}
    \centering
    \begin{tabular}{|c|c|c|c|c|c|c|c|}
        \hline
         ML Model & Feature Set & \# of Parameters & Accuracy & $TSS$ & $HSS_{2}$ & $F_1$ & $Bias$\\
        \hline
        \multirow{3}{*}{Logistic Regression} & SHARPs & 21 & 0.87 $\pm$ 0.01 & 0.79 $\pm$ 0.01 & 0.13 $\pm$ 0.02 & 0.15 $\pm$ 0.02 & 11.44 $\pm$ 1.75 \\
         & Topological & 21 & 0.87 $\pm$ 0.01 & 0.78 $\pm$ 0.02 & 0.12 $\pm$ 0.02 & 0.14 $\pm$ 0.02 & 11.88 $\pm$ 1.40 \\
         & Combined & 41 & 0.87 $\pm$ 0.02 & 0.79 $\pm$ 0.02 & 0.13 $\pm$ 0.02 & 0.15 $\pm$ 0.02 & 11.76 $\pm$ 1.98\\
        \hline
        \multirow{3}{*}{ERT} & SHARPs & 332 & 0.84 $\pm$ 0.01 & 0.79 $\pm$ 0.01 & 0.11 $\pm$ 0.01 & 0.13 $\pm$ 0.01 & 13.96 $\pm$ 0.81 \\
         & Topological & 483 & 0.85 $\pm$ 0.02 & 0.76 $\pm$ 0.04 & 0.11 $\pm$ 0.02 & 0.13 $\pm$ 0.02 & 13.34 $\pm$ 1.93 \\
         & Combined & 340 & 0.86 $\pm$ 0.02 & 0.77 $\pm$ 0.03 & 0.12 $\pm$ 0.02 & 0.14 $\pm$ 0.02 & 12.27 $\pm$ 2.12\\
        \hline
        \multirow{3}{*}{MLP} & SHARPs & 2198 & 0.85 $\pm$ 0.02 & 0.76 $\pm$ 0.02 & 0.11 $\pm$ 0.02 & 0.13 $\pm$ 0.02 & 13.13 $\pm$ 2.42 \\
         & Topological & 2198 & 0.85 $\pm$ 0.02 & 0.76 $\pm$ 0.02 & 0.11 $\pm$ 0.01 & 0.13 $\pm$ 0.01 & 13.56 $\pm$ 1.53 \\
         & Combined & 2918 & 0.86 $\pm$ 0.03 & 0.76 $\pm$ 0.03  & 0.11 $\pm$ 0.02 & 0.13 $\pm$ 0.02 & 13.10 $\pm$ 1.78 \\
        \hline
        \multirow{3}{*}{LSTM} & SHARPs & 6662 & 0.87 $\pm$ 0.02 & 0.75 $\pm$ 0.02 & 0.12 $\pm$ 0.02 & 0.14 $\pm$ 0.02 & 11.90 $\pm$ 1.93 \\
         & Topological & 6662 & 0.85 $\pm$ 0.02 & 0.75 $\pm$ 0.03 & 0.11 $\pm$ 0.02 & 0.13 $\pm$ 0.02 & 13.28 $\pm$ 2.11 \\
         & Combined & 7382 & 0.86 $\pm$ 0.01 & 0.75 $\pm$ 0.02 & 0.12 $\pm$ 0.01 & 0.14 $\pm$ 0.01 & 12.00 $\pm$ 1.63\\
        \hline
    \end{tabular}
    \caption{24-hour forecast performance of four
      machine-learning models using three feature sets. The
      third column shows the number of free parameters needed to
      classify a single data sample. For the ERT, this equals
      the average depth of the tree (the path taken by a data
      sample from the root to a leaf node in the tree). 
      }
    \label{tbl:models_features}
\end{table}
Note that the topological features perform just as well as the SHARPs
features. Moreover, the combined feature set does not provide any
improvement, confirming that
neither feature set provides a significant advantage over the other.
This is an answer to our second research question.  Abstract spatial
properties of active region magnetograms---calculated using abstract
universal algorithms that quantify shape from raw image data without
any assumptions about the underlying mechanics---appear to give ML
methods just as much traction on flare-forecasting problems as the set
of physics-based attributes that were hand-crafted by experts.

A between-model comparison addresses the first research question posed
in Section~\ref{sec:intro}: whether model complexity is an advantage
in the context of this problem.  The table suggests that the answer is
no.  Indeed, the general trend in the $TSS$ scores shows that
increased complexity slightly \textit{reduces} performance.  Indeed,
even though the MLP and LSTM models have orders of magnitude more
parameters, the simpler logistic regression and ERT models perform
marginally better, as judged by the $TSS$ scores.  (For the other
metrics, there is no significant variation among the four models.)
This may simply be due to data limitations; recall that the higher the
complexity of a machine-learning model, the more data is needed for
training.  If the training set is too small, the model will overfit
that data, causing it to fail to generalize well to the testing set.
Simpler models avoid this trap.  Note that while $\approx$ 460000
samples might seem large, many of the images in the solar flare data
set are similar, and almost all are non-flaring.  In other words, the
total amount of \textit{information}---i.e., the number of
sufficiently diverse and useful samples---is small in this data set.

Determining whether data limitations are in play is an important open
problem in current ML research.  One way to approach it is to compare
the values of the weighted binary cross-entropy loss function across
the models; another is to observe the patterns in the convergence over
the training process.  Both are problematic for the more-complex
models in our study.  The ERT model does not generate a loss, nor does
it have an iterative training process.  For models that do have an
iterative training procedure (MLP and LSTM), we carried out the second
test, and found that both the validation and training losses reached
asymptotes during the training process, suggesting---but of course not
proving---that overfitting is not at issue.  The first approach is not
useful in the case of the LSTM because one-to-one loss comparisons are
problematic in such models due to the variation in the number of
samples in the training and testing sets that occurs when the original
data are converted to temporal sequences by the feedback loop in the
model.  In the future, as more data is recorded by SDO/HMI, we hope to
learn, by re-running these experiments on longer, richer data sets,
whether the effects described in this paragraph are artifacts of data
limitations or something more fundamental.

To assess the effects of the prediction horizon,
we carried out a set of experiments using the best-performing
model/feature combination from Table~\ref{tbl:models_features}
(logistic regression with the combined feature set) and generated
forecasts for 3, 6, and 12 hours in addition to our previous 24 hour
case.  The results are shown in Table~\ref{tbl:forecasting_window}.
As one would expect, the shorter the forecast window, the higher the
accuracy, but the story for the other metrics is more complicated:
$TSS$ is roughly similar for all forecasting windows, while $HSS_{2}$,
$F_1$, and $Bias$ actually \textit{worsen} as the forecast window
shrinks.  This counterintuitive result is almost certainly due to the
increasing fraction of negative samples: for each flare, fewer
magnetograms will be labeled as ``flaring within $m$ hours'' if $m$ is
smaller.  This is, again, a side effect of tuning on the $TSS$ score,
as discussed above: its optimization leads to a high false positive
rate for a severely imbalanced dataset.
\begin{table}

    \begin{tabular}{|c|c|c|c|c|c|}
        \hline
         Forecasting Window & Accuracy & $TSS$ & $HSS_{2}$ & $F_1$ & $Bias$ \\
         \hline
         24 hours & 0.87 $\pm$ 0.01 & 0.79 $\pm$ 0.01 & 0.13 $\pm$ 0.02 & 0.15 $\pm$ 0.02 & 11.44 $\pm$ 1.75 \\
         12 hours & 0.89 $\pm$ 0.02 & 0.82 $\pm$ 0.01 & 0.10 $\pm$ 0.03 & 0.11 $\pm$ 0.03 & 17.13 $\pm$ 4.10\\
         6 hours & 0.91 $\pm$ 0.01 & 0.82 $\pm$ 0.02 & 0.07 $\pm$ 0.02 & 0.07 $\pm$ 0.02 & 25.17 $\pm$ 4.83 \\
         3 hours & 0.93 $\pm$ 0.01 & 0.79 $\pm$ 0.05 & 0.05 $\pm$ 0.01 & 0.05 $\pm$ 0.01 & 33.18 $\pm$ 4.60 \\
         \hline
    \end{tabular}

    \caption{Logistic regression forecasts for different horizons.}
    \label{tbl:forecasting_window}
\end{table}

The LSTM is not only the most complicated model in this study---by a
factor of three, as judged by the number of free parameters---but also
the only one that uses the AR \textit{history}.  In view of this, its
lack of performance is particularly striking.  The feedback loop in
this architecture makes it difficult to deconvolve the effects of the
temporal history and the number of parameters, so we cannot say for
sure whether or not the former, alone, confers any advantage.
Nevertheless, the additional free parameters certainly do not appear
to help.

\begin{figure}
	\centering
	\gridline{
		\fig{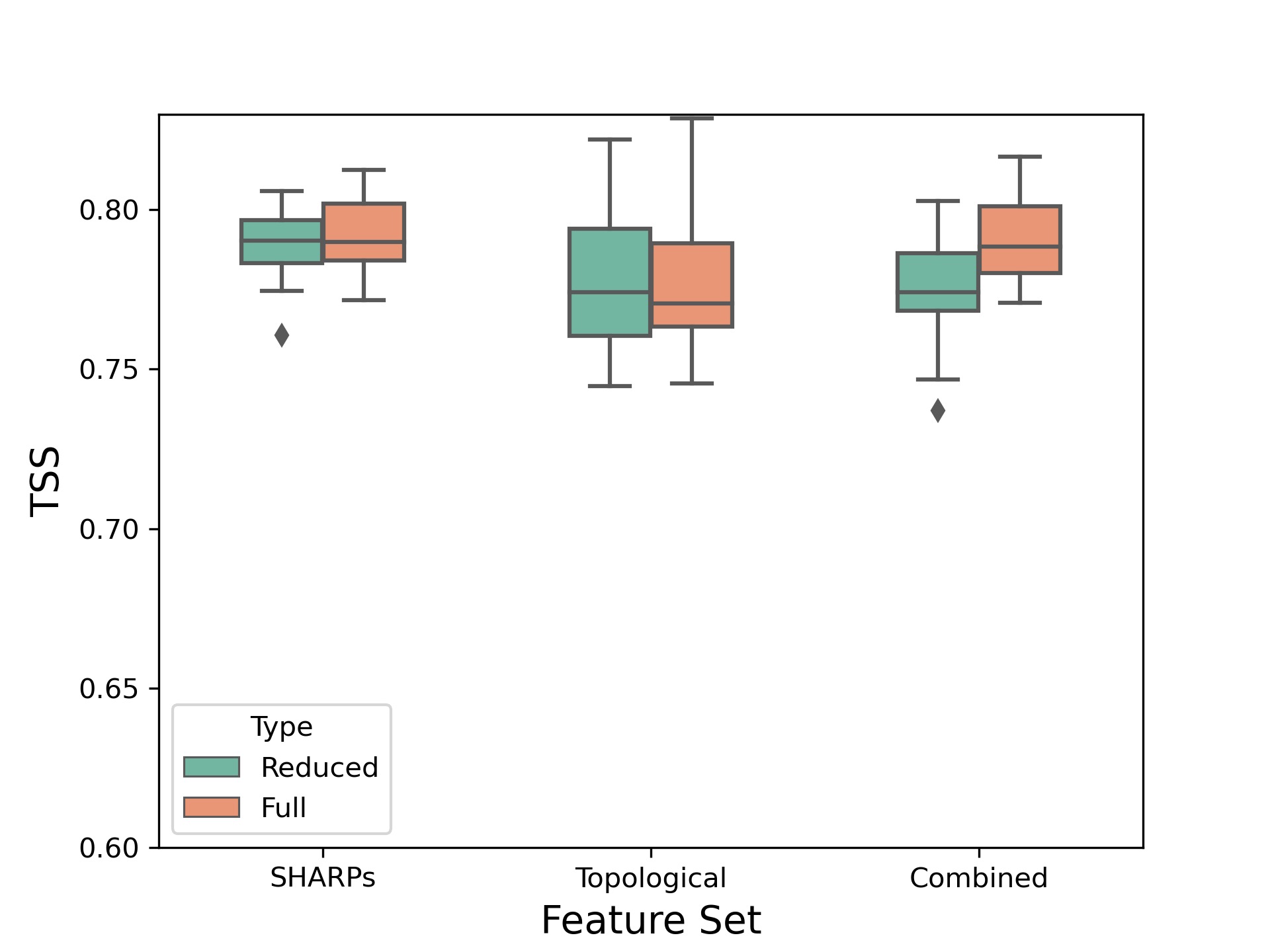}{0.4\textwidth}{(a) Logistic Regression}
		\fig{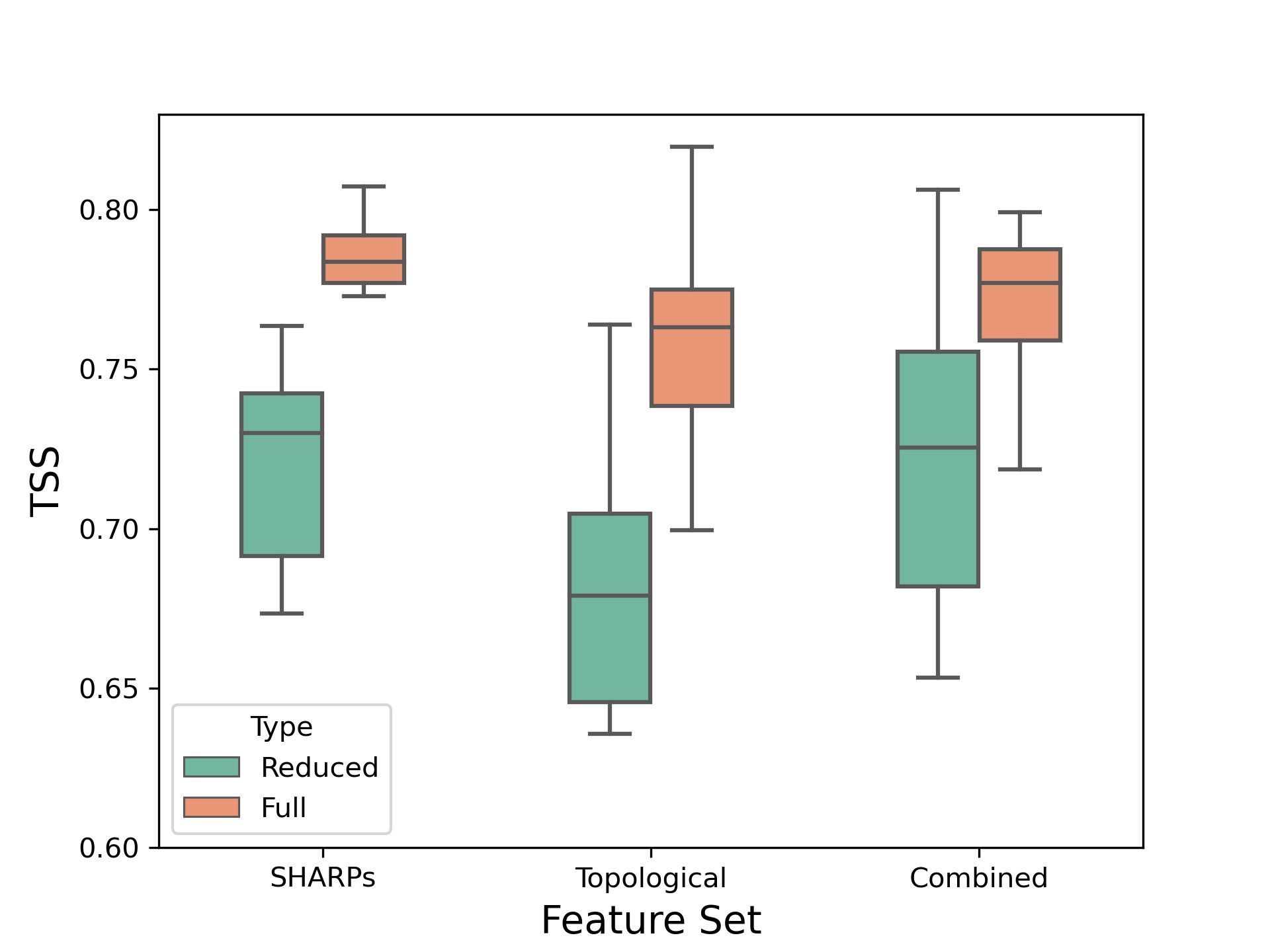}{0.4\textwidth}{(b) ERT}
	}
	\gridline{
		\fig{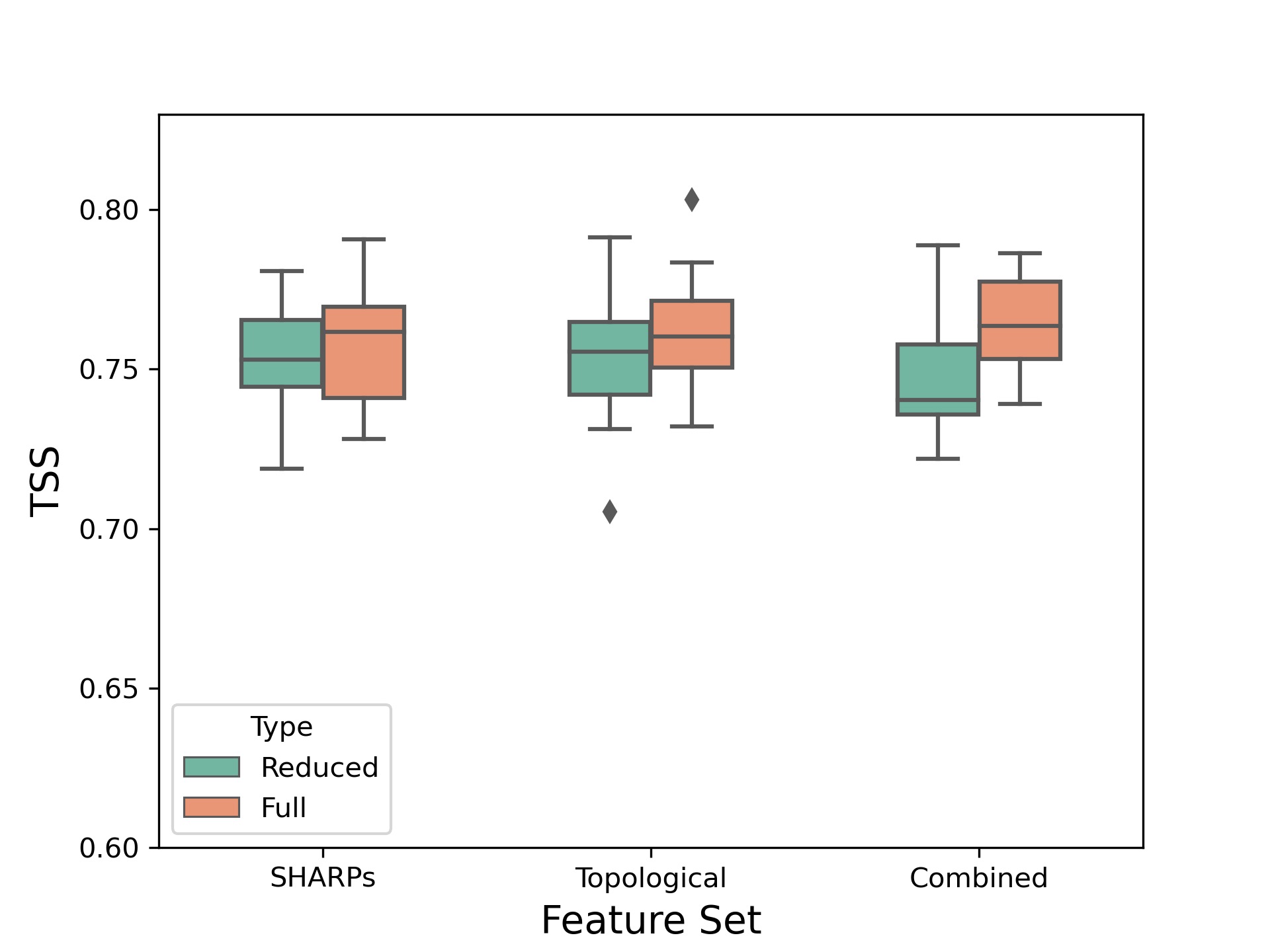}{0.4\textwidth}{(c) MLP}
		\fig{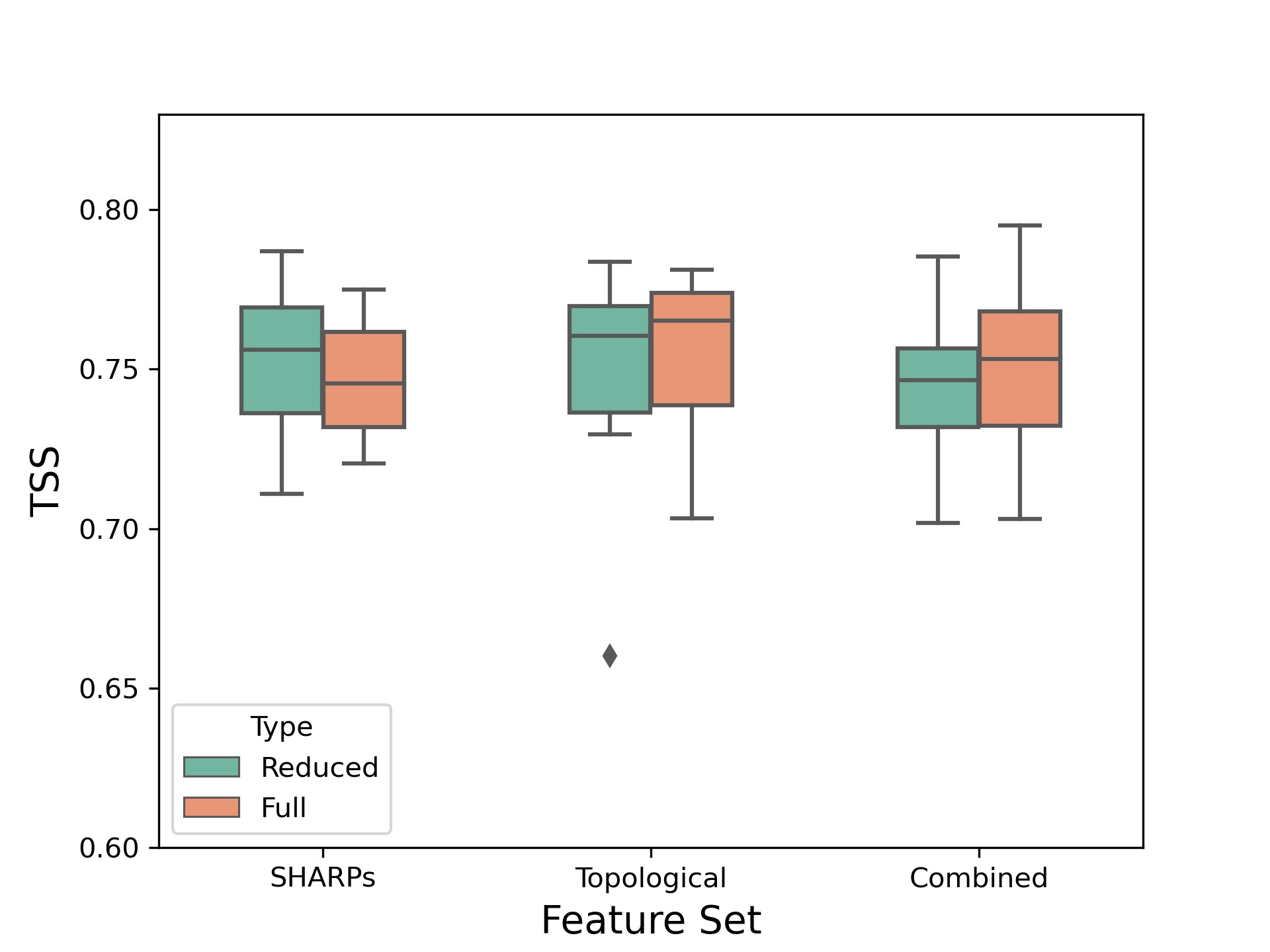}{0.4\textwidth}{(c) LSTM}
	}
	\caption{$TSS$ score comparison in the form of box-whiskers plots 
	for four ML models, trained on the three feature sets and 
	their PCA reduced counterparts over ten trials. The central line
	each box represents the median $TSS$ score while the top and 
	bottom edge of the boxes represent the 25 and 75 percentiles
	over the ten trials. The whiskers on either sides are the upper and
	lower bounds for the scores in each experiment, and the dots
	represent the outliers (see the plotting manual \url{https://seaborn.pydata.org/generated/seaborn.boxplot.html}
	for the definition of outliers). After PCA reduction, the SHARPs
	feature set is reduced from 20 to 9 features, the
	topological set from 20 to 3 features, and the combined set
	from 40 to 11 features. For all models except for the ERT, 
    the reduced feature set does just as well as the complete 
    feature set.}
\label{fig:lr_pca_stats}
\end{figure}

Our final research question concerned reduction of the dimension of
the feature sets. To explore this,
we re-process the data sets to calculate values for the PCA-based
features detailed in Section~\ref{sec:pca}, using 9, 3, and 11 PCA
vectors for the traditional, topological and combined feature sets,
respectively.  We then re-tune the model hyperparameters and repeat
the train/test procedure.  The results are shown in
Fig.~\ref{fig:lr_pca_stats}.
For all three feature sets, the performance of the full and
reduced-order models are similar for the logistic regression, MLP, and
LSTM models, as judged by $TSS$ scores.  This extends to the other
metrics as well (not shown).  In other words, we can successfully
simplify these ML models by reducing the number of
features \textit{without sacrificing performance}.  In view of the
discussion above about data limitations, this is an obvious advantage,
as models that work with smaller feature sets have fewer free
parameters that must be learned from the same training data.

The ERT model, however, departs from this pattern, demonstrating
a marked reduction in $TSS$ scores for the PCA-reduced feature set.
This effect is strongest for the topological feature set, followed by
the SHARPs feature set, with the combined feature set seeing the
smallest impact.  A possible explanation for this performance
degradation lies in the way ERTs are
constructed.  In a $k$-feature ERT, $\sqrt{k}$ randomly chosen
features are typically used to  determine the split at each
branch point.  A reduction in the number of features, then,
confines the exploration of the ERT parameter space.  This could result in
branches that do not effectively separate the positive and negative
samples, thereby impacting the model performance.
Modifying the training process to use all $k$ features for choosing
the best split can ameliorate this problem.  Indeed after doing this,
we find that there is no statistically significant difference between
the reduced and the full feature sets for the $TSS$ scores of the
topological and combined feature sets.
However, the SHARPs feature set does not show a similar improvement
upon increasing the number of features used in the split.  This
disparity suggests that the presence of shape-based features in the
combined set
helps make up for the
shortcomings of the physics-based features.  This agrees with the
results in
Section~\ref{sec:pca} that show both SHARPs and topological features are
highly weighted in the first principal component of the combined
feature set.

\section{Conclusions} \label{sec:conclusion}

Machine learning-based solar flare prediction has been a topic of
interest to the space weather community for some time.  Various
machine-learning models, ranging from simple tools like logistic
regression to incredibly complex ``deep-learning'' models such as
multi-layer perceptrons (MLPs) and long short-term memories (LSTMs),
have been used to map the correlation between physics-based magnetic
field features and the flaring probability in the near future.  There
have been studies that compared different machine-learning
flare-forecasting models \citep{barnes_survey, Nishizuka:2017,
  Florios2018}; however, the dominant focus of these papers has been
on performance comparison and they only used the basic SHARPs feature
set.  Furthermore, none of these approaches performed a systematic,
automated hyperparameter tuning process to assure a fair comparison,
and only \cite{Florios2018} did any hyperparameter tuning at all
(using a grid search method on discrete hand-selected values).

Our first objective in this paper was to systematically compare a set
of machine-learning models and determine whether higher complexity is
correlated with better performance.  Using an automated hyperparameter
tuning approach, we compared four models with increasing
complexity: logistic regression, extremely randomized trees, MLP and
LSTM. Across multiple experiments, and using an automated
hyperparameter tuning approach, we showed that increasing the
complexity of the model did not improve the model performance for the
24-hour M1.0+ flare-forecasting problem.  (Note that we do not include
convolutional neural networks as part of this study due to their high
computational needs.  However, we do address them separately in
\cite{Deshmukh2021Arxiv}.)

Secondly, we evaluated a new shape-based feature set that was introduced
in \cite{Deshmukh2020}m composed of quantities extracted from
magnetograms using topological data analysis.
The results from these features are compared
to the standard physics-based SHARPs feature set.  Our results extend
previous pilot studies \citep{Deshmukh2020,
Deshmukh2020_IAAI, Knyazeva2017} by using a far more comprehensive
SHARPs feature set, employing four different ML models, and using a
systematic hyperparameter tuning methodology to ensure fairness in
comparison.  This broader and deeper study confirmed that the
shape-based features---calculated using abstract and universal algorithms
without any assumptions about the underlying physics---provide as
much traction to ML models as the set of physics-based attributes that
were hand-crafted by experts.  Further, the combination of the
shape-based and physics-based feature sets does not provide any
improvement over the individual sets, confirming that neither feature
set provides a significant advantage over the other.  

Lastly, we studied a different aspect of model complexity: the
dimensionality of the feature set.  Using principal component analysis
(PCA) to find relevant subspaces of the feature space, we studied the
effect of dimensionality on model performance.  We found that the
topological feature set afforded the largest dimensionality
reduction, and that the PCA-reduced datasets performed just as well
as the original feature sets for three of the four ML
models.\footnote{The exception, the ERT model, did not improve due to
architectural design choices.}  In a data-limited situation, this is a
major advantage for the effectiveness of machine-learning based
solar-flare prediction methods, since more-complex models generally
require larger training sets.

\section*{Acknowledgments}

This study was funded by a grant from the National Science Foundation (Grant No. AGS 2001670) and by a grant from the NASA Space Weather Science Applications Program (Grant No. 80NSSC20K1404).

\newpage
\bibliography{references}{} 
\bibliographystyle{aasjournal}



\end{document}